\documentclass[twocolumn, twocolappendix]{aastex631}
\usepackage{amsmath}
\usepackage{booktabs}
\usepackage{CJK}
\usepackage[bbgreekl]{mathbbol}
\usepackage{bbold}
\newcommand{\beq}{\begin{equation}}
\newcommand{\baln}{\begin{aligned}}
\newcommand{\eeq}{\end{equation}}
\newcommand{\ealn}{\end{aligned}}
\newcommand{\beqn}{\begin{eqnarray}}
\newcommand{\eeqn}{\end{eqnarray}}

\newcommand{\Alfven}{Alfv\'{e}n }

\newcommand{\earthpotential}{\Phi_{\rm 1au}}

\shortauthors{Li et al.}
\shorttitle{Improved Solar Modulation Model for the Inner Heliosphere}

\begin{document}
\begin{CJK*}{UTF8}{gbsn}

\title{
Galactic Cosmic-Ray Propagation in the Inner Heliosphere: Improved Force-field Model
}


\author[0000-0003-1671-3171]{Jung-Tsung Li (李融宗)}
\affiliation{Center for Cosmology and AstroParticle Physics, The Ohio State University, Columbus, OH 43210, USA}
\affiliation{Department of Physics, The Ohio State University, Columbus, OH 43210, USA}
\affiliation{Department of Astronomy, The Ohio State University, Columbus, OH 43210, USA}

\author[0000-0002-0005-2631]{John F.~Beacom}
\affiliation{Center for Cosmology and AstroParticle Physics, The Ohio State University, Columbus, OH 43210, USA}
\affiliation{Department of Physics, The Ohio State University, Columbus, OH 43210, USA}
\affiliation{Department of Astronomy, The Ohio State University, Columbus, OH 43210, USA}

\author[0000-0002-8040-6785]{Annika H.~G.~Peter}
\affiliation{Center for Cosmology and AstroParticle Physics, The Ohio State University, Columbus, OH 43210, USA}
\affiliation{Department of Physics, The Ohio State University, Columbus, OH 43210, USA}
\affiliation{Department of Astronomy, The Ohio State University, Columbus, OH 43210, USA}

\email{li.12638@osu.edu, beacom.7@osu.edu, peter.33@osu.edu}

\date{\today}


\begin{abstract}
A key goal of heliophysics is to understand how cosmic rays propagate in the solar system's complex, dynamic environment. One observable is solar modulation, i.e., how the flux and spectrum of cosmic rays change as they propagate inward. We construct an improved force-field model, taking advantage of new measurements of magnetic power spectral density by Parker Solar Probe to predict solar modulation within the Earth's orbit. We find that modulation of cosmic rays between the Earth and Sun is modest, at least at solar minimum and in the ecliptic plane. Our results agree much better with the limited data on cosmic-ray radial gradients within Earth's orbit than past treatments of the force-field model. Our predictions can be tested with forthcoming direct cosmic-ray measurements in the inner heliosphere by Parker Solar Probe and Solar Orbiter. They are also important for interpreting the gamma-ray emission from the Sun due to scattering of cosmic rays with solar matter and photons.
\end{abstract}


\section{Introduction}

How do charged cosmic rays propagate in dynamic magnetic environments? Despite decades of work, the uncertainties remain large. This is true even for the solar system, where we have rich auxiliary data. Getting good agreement between theory and observation here is a prerequisite for understanding more distant astrophysical systems. It will also lead to better probes of the Sun's magnetic fields and how those and cosmic rays affect Earth, spacecraft, and the solar system itself.

In the solar system, as galactic cosmic rays (GCRs) diffuse inward, they are increasingly affected by solar modulation, which reduces their energy and intensity. This is caused by the GCRs undergoing interactions with magnetic fluctuations in the solar wind as the wind convects outward. The strength of the modulation varies over the solar cycle, being less at solar minimum. The foundational data for modulation studies is the collection of GCR spectra at Earth~\citep{1998ApJ...502..278A, 2000ApJ...532..653B, 2000ApJ...545.1135S, 2013ApJ...765...91A, 2014PhRvL.113l1102A, 2015ApJ...810..142A, 2015PhRvL.114q1103A, 2015PhRvL.115u1101A, 2016ApJ...822...65A, 2017PhRvD..95h2007A, 2017Natur.552...63D, 2020ApJ...893..145M, 2021PhRvL.127A1102A, 2022PhRvL.128w1102A}, which are well measured for many species over wide energy ranges. To determine the incoming GCR spectra, these data are compared to less precise but crucial measurements made throughout the outer solar system, including with Voyager~1 and~{2} out to the heliospheric boundary at $\approx 100$~au~\citep{2013GeoRL..40.1665W, 2013Sci...341.1489G, 2013Sci...341..150S,  2016ApJ...831...18C}.

There are new opportunities to probe GCRs at the opposite extreme: the inner solar system, which has barely been explored. \emph{Direct} probes of GCR spectra will soon be provided by the US-led Parker Solar Probe (PSP;~\cite{2016SSRv..204....7F}), which will reach $<10~r_\odot$ ($<0.047$~au) and the European-led Solar Orbiter (SolO;~\cite{2020A&A...642A...1M}), which will reach $60~r_\odot$ (0.279~au). Together, they will probe energies of $\simeq$~{10--200}~MeV for hadronic GCRs~\citep{2016SSRv..204..187M, 2021A&A...656A..22W}.

\emph{Indirect} probes are provided by the gamma rays produced by GCR interactions with the Sun~\citep{1991ApJ...382..652S, 2006ApJ...652L..65M, 2007Ap&SS.309..359O, 2008A&A...480..847O, 2021JCAP...04..004O, 2011ApJ...734..116A, 2016PhRvD..94b3004N, 2017PhRvD..96b3015Z, 2018PhRvD..98f3019T, 2018PhRvL.121m1103L, 2022PhRvD.105f3013L, 2020arXiv200903888L, 2020PhRvD.101h3011M}. The dominant emission from the solar disk ($\theta < 0.{}^\circ 25$) is caused by hadronic GCR interacting with matter in the photosphere. The dominant emission from the solar halo ($\theta \lesssim 20^\circ$) is caused by electron GCRs interacting with photons out to $\simeq 1$~au from the Sun. These data are sensitive to GCRs over a wide range (so far, $\simeq$ 1--1000~GeV), but what they reveal about the Sun is clouded by uncertainties about their modulation in the inner solar system.

Predictions of the GCR spectra in the inner heliosphere are thus urgently needed. Most of the numerical solutions of Parker's transport equation focus on reproducing GCR spectra at Earth's orbit and the outer solar system~\citep{2013AdAst2013E...1B, 2017ApJ...846...56Q, 2018AdSpR..62.2859B, 2019ApJ...873...70A, 2021ApJ...909..215A, 2019ApJ...878...59B, 2020ApJ...894..121M}. The little work that has been done for the inner solar system uses the force-field model to calculate solar modulation~\citep{2006ApJ...652L..65M, 2008A&A...480..847O, 2011ApJ...734..116A, 2022PhRvD.105f3013L}. The force-field model is a one-dimensional diffusion-convection equation in the stationary solar system frame~\citep{1967ApJ...149L.115G, 1968ApJ...154.1011G}. It is parameterized by the force-field modulation potential energy $\Phi$, with higher $\Phi$ corresponding to stronger modulation. It is often assumed that the mean free path is linearly proportional to particle rigidity, which leads to the modulation potential energy being rigidity independent. Under this form of modulation potential energy, it has been shown that the force-field model can simply quantify the variability of solar modulation in time, which has practical applications in studies of atmospheric ionization, radiation environment, and radionuclide production~\citep{2005JGRA..11012108U, 2017JGRA..122.3875U}. A particular rigidity-dependent form of the modulation potential energy was discussed in~\cite{1971Ap&SS..11..288C}, \cite{1972Ap&SS..16...55U}, \cite{1972Ap&SS..17..426U}, and \cite{1973Ap&SS..20..177U} for evaluating the solar modulation of low-energy GCRs in the outer solar system.

However, the force-field model is known to have shortcomings. For example, three-dimensional particle drifts and heliospheric current sheets are not considered, limiting the model's ability to determine the latitudinal dependence of the GCR distribution. In addition, a force-field model does not provide good predictions for the GCR spectra in the outer heliosphere~\citep{2004JGRA..109.1101C}. Ultimately, full numerical calculations of the cosmic-ray transport equation are required. Until then, approximations are needed for rapid, accessible use.

In this paper, we develop an improved version of the force-field model, now taking into account the radial evolution of the turbulence in the inner heliosphere and the rigidity dependence of the modulation potential energy. This is made possible by using PSP magnetometer data that reveal the power spectral density (PSD) of magnetic fluctuations, which have been measured down to 0.17~au for the first time~\citep{2020ApJS..246...53C}. In short, the behavior of the magnetic PSD measured by PSP is different from naive considerations in which the modulation potential energy is rigidity independent. Ultimately, we find that GCR modulation in the inner solar system is very modest, i.e., that the spectra are close to those measured at Earth. This agrees with earlier hints from Helios, Pioneer, and~MESSENGER~\citep{1977ApJ...216..930M, 2016JGRA..121.7398L, 2019A&A...625A.153M}, which had limited data.

The rest of the paper is organized as follows. In Section~\ref{sec:FF_overview}, we review the theoretical framework for GCR propagation in the solar system. In Section~\ref{sec:diffusion}, we develop our new calculation of the diffusion coefficients and modulation potential energies for the inner solar system. In Section~\ref{sec:result}, we calculate the numerical results for the predicted GCR intensities and compare the calculated and measured GCR radial gradients. In Section~\ref{sec:conclusion}, we conclude and discuss the next steps.

\section{Overview of the Force-field Model}\label{sec:FF_overview}

In this section, we discuss the force-field model, its limitations, and paths to improvement. In Section~\ref{subsec:CR_transport}, we discuss the general case of cosmic-ray transport in interplanetary space. In Section~\ref{subsec:FF_formalism}, we review the derivation of the force-field solution and its associated characteristic equation. In Section~\ref{subsec:FF_R_indep}, we present the rigidity-independent modulation potential energy approach used in the solar gamma-ray literature. In Section~\ref{subsec:FF_R_dep}, we discuss why the \emph{rigidity-dependent} modulation potential energy leads to a more accurate force-field model.

\subsection{Cosmic-Ray Transport Equation}\label{subsec:CR_transport}

GCRs entering the solar system scatter from magnetic irregularities in the solar wind and random walk through interplanetary space as the wind expands outward from the Sun. In addition, cosmic rays experience gradient and curvature drifts due to the inhomogeneity of the large-scale interplanetary magnetic fields (IMF). The equation that describes cosmic-ray transport and solar modulation in the heliosphere is~\citep{1965P&SS...13....9P, 1978Ap&SS..58...21G}
\beq
\baln
    \frac{\partial U_p}{\partial t} &+ \boldsymbol{\nabla} \cdot\left(C \mathbf{V}_{\rm sw} U_p  \right) - \boldsymbol{\nabla} \cdot \left( \kappa \cdot \boldsymbol{\nabla} U_p \right) \\
    &+ \langle \mathbf{v}_D \rangle \cdot \boldsymbol{\nabla} U_p + \frac{1}{3}\frac{\partial}{\partial p} \left( p \mathbf{V}_{\rm sw} \cdot \boldsymbol{\nabla} U_p \right) = 0,
    \label{eq:full_transport_eq}
\ealn    
\eeq
where the frame of reference is fixed in the solar system. Here, $p$ is the particle momentum, $U_p$ is the differential number density of cosmic rays with respect to $p$, $\mathbf{V}_{\rm sw}$ is the solar wind velocity, $\kappa$ is the the second-rank diffusion tensor, $\langle \mathbf{v}_D \rangle$ is the drift velocity, and $C = 1 - \frac{1}{3U_p}\frac{\partial \left(p U_p\right)}{\partial p}$ is the Compton--Getting factor~\citep{1968Ap&SS...2..431G}. The differential number density $U_p$ is related to the differential intensity (flux per solid angle) $J_E$ (with respect to particle total energy $E$) by $U_p = 4 \pi J_E$.

There is a vast literature developing numerical models and solutions to the transport equation. The first three-dimensional transport calculation, including the effects of diffusion, drift, and heliospheric current sheets, was developed by~\cite{1983ApJ...265..573K}. The most recent developments include~\cite{2017ApJ...846...56Q}, who consider the diffusion coefficients from nonlinear guiding center theory as well as the latitudinal and radial dependence of magnetic turbulence. \cite{2018AdSpR..62.2859B} adopt a modified IMF in the polar region in their two-dimensional Monte Carlo code. \cite{2019ApJ...878...59B} consider the empirical models of diffusion and drift coefficients fitting the observed GCR spectra from PAMELA and Voyager~1. \cite{2019AdSpR..64.2459B} improve the accuracy of particle transport solutions during solar maxima by including the time dependence of the heliosphere boundary and heliosheath region.

In general, these numerical models provide state-of-the-art analyses. However, these models are not publicly available and are complicated to construct. In addition, predictions from these models for GCR spectra in the inner heliosphere (within 1~au from the Sun) are not yet available. Nevertheless, we consider their approach as a complete treatment for GCR modulation. Our improved force-field treatment is intended to provide a simple, yet reasonable, approximation for evaluating the inner heliosphere GCR intensity. We encourage new calculations with numerical models for the inner solar system.

\subsection{Force-field Model}\label{subsec:FF_formalism}

The force-field model derived by~\cite{1967ApJ...149L.115G, 1968ApJ...154.1011G} and \cite{1973Ap&SS..25..387G} is widely used due to its simplicity and inclusion of energy loss. It begins from the cosmic-ray transport equation in the solar system frame, as shown in Equation~\eqref{eq:full_transport_eq}, and assumes that (i) transport reaches a steady state, (ii) there is no source or sink of GCR in the heliosphere, (iii) propagation is spherically symmetric, and (iv) particle drift is not considered. The one-dimensional force-field equation is obtained by demanding the convection flux balance the diffusion flux in the radial direction, which yields
\beq
    \kappa_{rr} \frac{\partial U_p}{\partial r} + \frac{1}{3} V_{{\rm sw}} \: p^3 \frac{\partial}{\partial p}\left(\frac{U_p}{p^2}\right) = 0,
    \label{eq:force_field_eqn}
\eeq
where $\kappa_{rr}$ is the $\left(r,r\right)$ component of ${\kappa}$, which depends on $E$ and $r$. The solution of Equation~\eqref{eq:force_field_eqn} is a constant $U_p/p^2$ along a phase-space contour line following the characteristic equation $dp/dr = V_{{\rm sw}} p/3\kappa_{rr}$. Presented in terms of $E$ and $J_E$, the solution that connects the GCR intensity at the heliocentric radius $r_1$ to that at $r_2 > r_1$ is
\beq
    \frac{J_E\left(E, r_1\right)}{E^2-E_0^2} = \frac{J_E\left(E+\Delta\Phi, r_2\right)}{\left(E+\Delta\Phi\right)^2 - E_0^2},
    \label{eq:force_field_solution}
\eeq
where $\Delta \Phi$ is defined as the energy change of the particle from $r_1$ to $r_2$, obtained from the characteristic equation, which is
\beq
    \frac{dE}{dr} = \frac{ V_{{\rm sw}} }{ 3\kappa_{rr} }  \frac{\left(E^2 - E_0^2\right)}{E}  ,
    \label{eq:characteristic_eqn}
\eeq
with $E_0$ denoting the rest mass energy of the particle.

\subsection{Rigidity-independent Modulation Potential Energy}\label{subsec:FF_R_indep}

Previous work on the gamma-ray emission from the solar halo required the intensity of GCR electrons near the Sun~\citep{2006ApJ...652L..65M, 2008A&A...480..847O, 2021JCAP...04..004O, 2011ApJ...734..116A, 2022PhRvD.105f3013L}. They adopted the force-field model with an assumption that $\kappa_{rr}\sim R v \, r^\eta$, where $R$ is the particle rigidity, $v$ is the particle speed, and $\eta$ ranges from 1.1--1.4 for the entire heliosphere. Integrating Equation~\eqref{eq:characteristic_eqn}, they obtained a rigidity-independent form of $\Phi$ as 
\beq
    \Phi\left(r\right) = \frac{\lvert Z \rvert e}{3} \int^{ r_{\rm b} }_{r} \frac{V_{\rm sw}}{\kappa_{rr}/\left(Rv\right)} \: dr^\prime ,
    \label{eq:R_indep}
\eeq
where $\Phi$ is set to zero at the heliospheric boundary, $r_{\rm b}\approx 100 \: {\rm au}$. They expressed the rigidity-independent $\Phi$ in Equation~\eqref{eq:R_indep} as
\beq
    \Phi\left(r\right) = {\earthpotential}\: \frac{r^{1-\eta} - r_{\rm b}^{1-\eta}}{\left(1\:{\rm au}\right)^{\eta-1} - r_{\rm b}^{\eta-1}},
    \label{eq:R_indep_compact}
\eeq
assuming a constant $V_{\rm sw}$. The term $\earthpotential$ is the accumulated result of solar modulation from the heliospheric boundary to $1 \, {\rm au}$ which can be obtained from neutron monitor experiments, e.g,~\cite{2017JGRA..122.3875U}. In this work, $\Phi_{\rm 1au}$ is only used in the rigidity-independent case that we show as a comparison.

We emphasize that $\Phi$ in this work denotes \emph{modulation potential energy} experienced by nuclei. This is different from \emph{modulation potential} $\phi \equiv \Phi/\left( {|Z|e} \right) $ used in the literature, which denotes the potential experienced by each charged nucleon.

\subsection{Rigidity-dependent Modulation Potential Energy}\label{subsec:FF_R_dep}

The choice of $\kappa_{rr} \propto R$ in Equation~\eqref{eq:R_indep} could overestimate solar modulation effects. As we demonstrate in Section~\ref{sec:diffusion}, a more realistic magnetic turbulence condition would show that $\kappa_{rr}$ varies as $R^{1/3}$ to $R^{1/2}$ at low GCR energies and as $R$ at high energies. In particular, low-energy GCRs do not experience as much solar modulation as in the case in Section~\ref{subsec:FF_R_indep}, where $\kappa_{rr}$ is assumed to vary as $R$.

\section{Modulation potential energy in interplanetary space}\label{sec:diffusion}

In this section, we present our improved approach for evaluating solar modulation in the force-field model. In Section~\ref{subsec:kappa_rr}, we describe general role of the magnetic PSD in determining $\kappa_{rr}$ in the inner heliosphere. In Section~\ref{subsec:PSD}, we formulate the functional form of the magnetic PSD from the PSP measurements. In Section~\ref{subsec:parallel_diffusion}, we lay out the GCR diffusion model from the quasi-linear theory (QLT). In Section~\ref{subsec:parallel_diffusion_result}, we show the numerical results for the diffusion coefficients. In Section~\ref{subsec:modulation_potential}, we calculate the modulation potential energy from $\kappa_{rr}$. In the conclusion, we discuss the modulation potential energies associated with GCR propagation in the turbulent environment.

Limited by PSP's orbits and the operation time so far, we only consider the following conditions in our analysis: (1) GCR modulation in the solar ecliptic plane and (2) during the solar minimum. Furthermore, because we only take into account the IMF for the local mean magnetic field, our predictions go down only to 0.1~au. We do not consider the solar modulation in the coronal magnetic fields within $\sim 0.05$~au from the Sun.

\subsection{GCR Diffusion in the Inner Heliosphere}\label{subsec:kappa_rr}

Due to the large-scale IMF, $\kappa_{rr}$ is spatially anisotropic. Locally, the GCR diffusion coefficient is separated into components parallel ($\kappa_\parallel$) and perpendicular ($\kappa_\perp$) to the mean value of local IMF, respectively. At different locations in the solar ecliptic plane, $\kappa_{rr} = \kappa_{\parallel} \cos^2\psi + \kappa_{\perp} \sin^2\psi$, where $\psi$, known as the Parker spiral angle, is the angle between the IMF and the heliocentric radius vector. (The value of $\psi$ at $1 \; {\rm au}$ is $\sim 45^\circ$.) According to the numerical simulation in~\cite{1999ApJ...520..204G}, $\kappa_\perp/\kappa_\parallel \simeq 0.025$ for a representative of IMF. Using this value, it is easy to show that $\kappa_{\perp} \sin^2\psi > \kappa_{\parallel} \cos^2\psi$ for $\psi \gtrsim 81^\circ$ (or $r\gtrsim 5$~au). As a result, force-field modulation at the outer heliosphere is dominated by perpendicular diffusion, while in the inner heliosphere, it is dominated by parallel diffusion. Since we focus on modulation within 1~au where perpendicular diffusion is negligible, we neglect $\kappa_\perp \sin^2 \psi$ and approximate $\kappa_{rr}$ as $\kappa_{\parallel} \cos^2\psi$.

To calculate $\kappa_\parallel$, we need to know the PSD of the magnetic fluctuations (we show the formalism in Section~\ref{subsec:parallel_diffusion}). This is because diffusion parallel to the mean magnetic field is caused by GCR resonant interactions with magnetic fluctuations. The stronger the magnetic fluctuations, the harder it is for particles to diffuse, and hence the lower $\kappa_\parallel$ is. For this purpose, we adopt PSP's measurement from~\cite{2020ApJS..246...53C} to describe the magnetic PSD in the inner heliosphere solar wind, as shown in Section~\ref{subsec:PSD}.

\subsection{Turbulence Spectrum}\label{subsec:PSD}

\begin{figure}[t]
    \centering
    \includegraphics[width=0.48\textwidth]{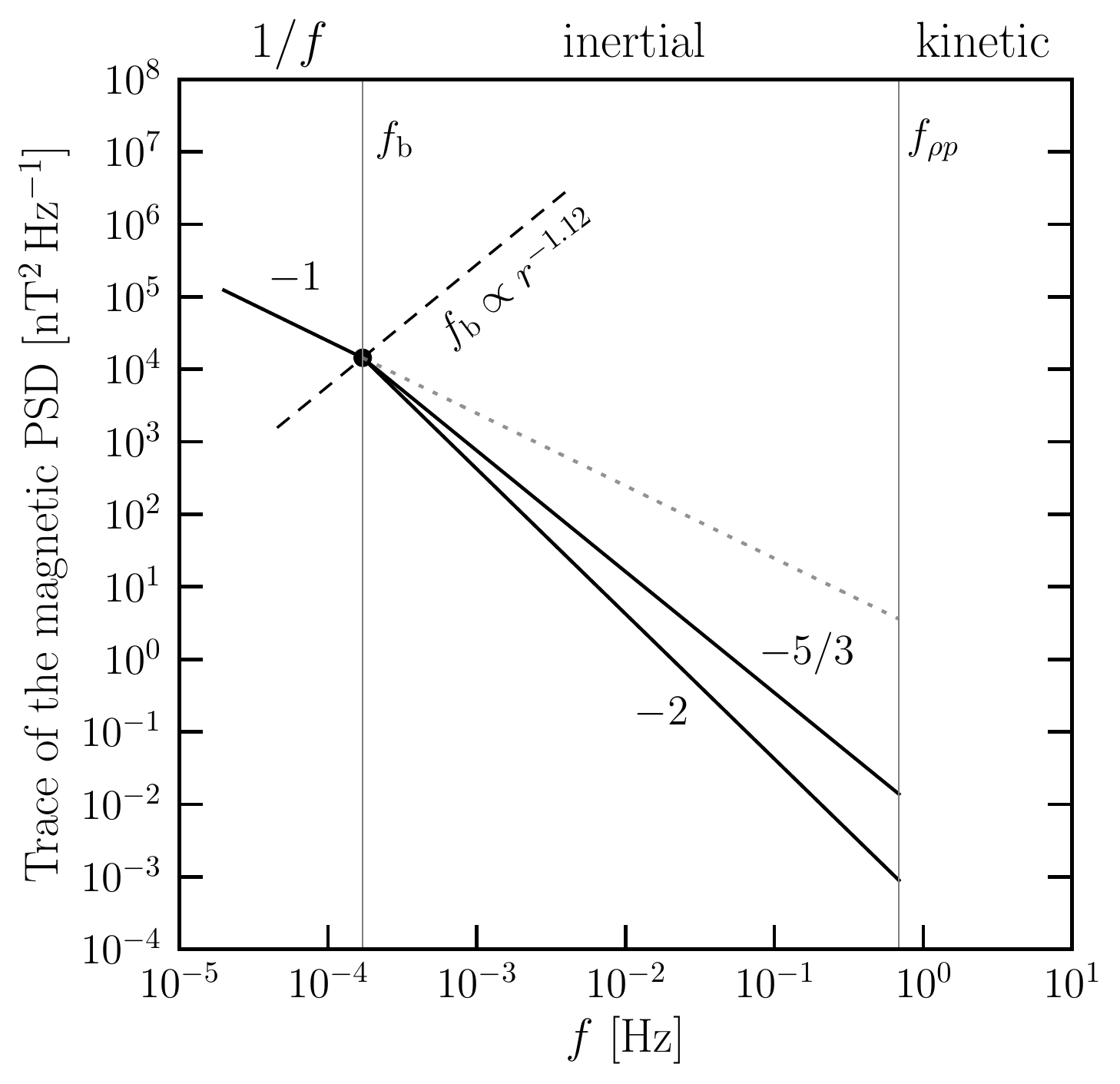} 
    \caption{Schematic diagram of the trace magnetic PSD and frequency break at 0.82~au. In the inertial range, the trace PSD is dominated by the perpendicular turbulence, which scales as $f^{-5/3}$. The black solid circle is where $f_{\rm b}$ is at $r=0.82~{\rm au}$. The black dashed line shows the scaling of $f_{\rm b}$ as $r$. For reference, we also draw the PSD of the parallel turbulence, which scales as $f^{-2}$. The gray dotted line is the extension of the $1/f$ power law to higher frequencies to contrast the change in the inertial range.}
   \label{fig:schematics}
\end{figure}

The observed temporal PSD of the magnetic fluctuations for frequency $f$ and at a heliocentric distance $r$ in the solar ecliptic plane is defined as the Fourier transform of the two-time correlation of the fluctuating magnetic fields $\delta \mathbf{B}$. It is expressed in tensor form as
\beq
    P_{{\rm B},{ij}}\left(f, r\right) = \int_{-\infty}^{\infty} \langle \delta {B}_i\left(r, t\right) \delta {B}_j\left(r, t+\tau\right) \rangle e^{-2\pi i f \tau} d\tau,
\eeq
where $t$ and $t+\tau$ refer to two different times. (The connection between $P_{{\rm B},{ij}}$ and $\kappa_{rr}$ will be given in the next subsection.) The matrix trace of $P_{{\rm B},{ij}}\left(f, r\right)$ within 1~au is theoretically expected to show three distinct power laws. The low-frequency region, called the ``$1/f$ range,'' is an $f^{-1}$ power law. The middle-frequency region, called the ``inertial range,'' is an $f^{-3/2}$ to $f^{-5/3}$ power law, depending on the magnetic conditions. The high-frequency region, called the ``kinetic range,'' varies as $f^{-2.3}$ or steeper. These three power laws have been identified in earlier data from Ulysses, Helios~I, Wind, and MESSENGER down to 0.3~au~\citep{2013LRSP...10....2B, 2015ApJ...805...46T}, but PSP will provide precise measurements of the changes in spectral shape and total energy of PSD along its trajectory down to 0.047~au.

We adopt the early PSP measurements from~\cite{2020ApJS..246...53C} to describe the observed PSD of the magnetic fluctuations in the solar wind. Their data were taken from 2018 October 6 to 2019 April 18, corresponding to the solar minimum at the end of the Solar Cycle 24. The heliocentric distance that PSP traveled during this period ranged from 0.17--0.82~au.

Figure~\ref{fig:schematics} illustrates their key findings about the spectrum. They are (i) the magnitude of matrix trace of the PSD, (ii) an $f^{-1}$ power law from $2 \times 10^{-5} \, {\rm Hz}$ up to the frequency break $f_{\rm b}$, (iii) an $f_{\rm b}$ that shifts to lower frequencies at greater heliocentric distances, (iv) the spectral shape of the inertial-range turbulence at $10^{-3}\,{\rm Hz}\lesssim f \lesssim 10^{-1}\,{\rm Hz}$, and (v) an $f^{-3/2}$ power law for the trace PSD at $r \approx 0.17\:{\rm au}$ and a gradual transition to $f^{-5/3}$ at $r \approx 0.6\:{\rm au}$ and beyond.

To simplify the analysis of the force-field model, we formulate the functional forms of the magnetic PSD from~\cite{2020ApJS..246...53C} with additional approximations and assumptions.
\begin{enumerate}
\item For the $1/f$ range, we approximate their trace PSD at $r=0.82\:{\rm au}$ as approximately $2.46 \: {\rm nT^2}/f$. The turbulence evolution in this range is described by the WKB approximation and scales as $r^{-3}$~\citep{1973ApJ...181..547H, 1982JGR....87.3617B, 1990JGR....9511945M}. This allows us to express the trace PSD in this range as
\beq
    \sum_{\rm i=j} P_{{\rm B}, ij} \left(f, r\right) = \frac{2.46 \: {\rm nT^2}}{f} \times \left( \frac{r}{\rm 0.82~au} \right)^{-3},
    \label{eq:E_B}
\eeq
where ${i=\emph{R,T,N}}$ are the solar ecliptic coordinates, with \emph{R} in the direction outward from the Sun in the solar ecliptic plane, \emph{N} normal to that plane, and \emph{T} perpendicular to \emph{R} and \emph{N}. In this plane, \emph{N} is normal to the mean magnetic field $\langle \mathbf{B} \rangle$. (Here to obtain $2.46 \: {\rm nT^2}/f$, we digitize the PSD plot in their Figure~1. Because the curves in the $1/f$ range fluctuate a lot, we picked the mean value of the curves.)
\item For the frequency break $f_{\rm b}$, we deduce it from their radial dependence of the break timescale $\tau_{\rm b}\propto r^{1.12}$ with $\tau_{\rm b} \approx 4.71\times 10^{3}\:{\rm s}$ at $r = 0.68 \: {\rm au}$. This allows us to write the functional form of $f_{\rm b}\equiv 1/\tau_{\rm b}$ separating the $1/f$ range and the inertial range as
\beq
    f_{\rm b} = 2.12\times 10^{-4}~{\rm Hz} \times \left(\frac{r}{\rm 0.68\,{\rm au}}\right)^{-1.12}.
    \label{eq:frequency_break}
\eeq
\item For the inertial range, the magnetic field spectral index is $\nu\approx -3/2$ at $r=0.17~{\rm au}$ and gradually shifts to $\nu\approx -5/3$ at $r=0.6~{\rm au}$. We approximate their result of $\nu$ as linear to $\ln r$ and thus express $\nu$ as 
\beq
    \nu\left(r\right) \approx - 1.5 - 0.135 \, \ln\left(\frac{r}{\rm 0.17~au}\right),
    \label{eq:spectral_index}
\eeq
for $0.17~{\rm au} \leq r \leq 0.6~{\rm au}$. Within $0.17~{\rm au}$ and beyond $0.6~{\rm au}$, we assume $\nu$ to be $-3/2$ and $-5/3$, respectively.
\end{enumerate}

In addition to the PSP measurements from~\cite{2020ApJS..246...53C}, we use three other earlier results. First, \cite{2010MNRAS.407L..31W} analyzed the low-frequency turbulence of the solar wind from Ulysses data and showed that the $1/f$ fluctuations are nearly isotropic. This finding allows us to approximate  $P_{{\rm B,NN}}\left(f, r\right)$ in the $1/f$ range as
\beq
    P_{\rm B, NN}\left(f, r\right) = \frac{1}{3} \sum_{\rm i=j} P_{{\rm B}, ij}\left(f, r\right).
\eeq
For $f > f_b$, the power law of $P_{{\rm B,NN}}\left(f, r\right)$ follows the spectral-index approximation in Equation~\eqref{eq:spectral_index}. As discussed in Section~\ref{subsec:parallel_diffusion}, $P_{{\rm B,NN}}\left(f, r\right)$ governs the diffusivity of GCRs parallel to $\langle \mathbf{B} \rangle$.

Second, \cite{2009PhRvL.102w1102S} analyzed the high-frequency turbulence of the solar wind from Cluster data and showed that the inertial range terminates at the Doppler-shifted proton gyroscale, $f_{\rho p} \equiv V_{\rm sw}/\left(2 \pi \rho_p\right)$, with $\rho_p$ as the thermal proton gyroradius of the solar wind. Above $f_{\rho p}$, MHD turbulence enters the kinetic range, at which the magnetic power spectra vary as $f^{-2.3}$ or steeper. Because of the weak magnetic power, we neglect scattering in the kinetic range. To calculate $\rho_p$ at different $r$ in the inner heliosphere, we adopt the analytical fit of the thermal proton temperature of the solar wind from~\cite{2009ApJ...702.1604C}.

Third, we emphasize that the $f^{-1}$ power law does not extend to arbitrarily low frequencies. \cite{2020ApJS..246...53C} provide a PSD with an $f^{-1}$ power law down to $2\times 10^{-5}~{\rm Hz}$. \cite{1986PhRvL..57..495M} have shown that this scaling at 1~au only extends to $\sim 2 \times 10^{-6} \, {\rm Hz}$, below which the spectra become flatter than $f^{-1}$. However, there are not enough data for us to properly model the PSD below this frequency range. We thus do not consider the GCR interactions with waves in this frequency range. As a result, we assume that the trace PSD in the $1/f$ range in Equation~\eqref{eq:E_B} is valid for $f > 2 \times 10^{-6} \, {\rm Hz}$. Within 1~au from the Sun and during the low solar activity cycle, magnetic fluctuations of $f\gtrsim 2 \times 10^{-6} \, {\rm Hz}$ resonantly interact with particles of $T\lesssim 40\,{\rm GeV}$. Therefore, our analysis is strictly only valid for $T\lesssim 40\,{\rm GeV}$. At higher energies, however, the modulation is negligible anyway.

\subsection{Parallel Diffusion Model}\label{subsec:parallel_diffusion}

In a weak turbulent plasma where particle relaxation is slow compared to particle gyration, the mean evolution of the particle distribution along the mean magnetic field can be described by QLT. According to this theory, the relation between the spatial diffusion coefficient parallel to the mean magnetic field, $\kappa_\parallel$, and the pitch-angle diffusion coefficient, $D_{\mu \mu}$, in a magnetostatic, dissipationless turbulence with slab geometry is given by~\citep{1966ApJ...146..480J, 1968ApJ...152..671J, 1975RvGSP..13..547V, 1976JGR....81.2089L}
\beq
    \kappa_\parallel \left(v, r\right) = \frac{v^2}{4} \int_{\mu_{{\rm min},s}}^{1} \frac{\left(1 - \mu^2\right)^2}{D_{\mu \mu}} d\mu,
    \label{eq:diffusion_QLT}
\eeq
where 
\beq
    D_{\mu \mu} = \frac{1 - \mu^2}{2 \lvert \mu \rvert v} \left(\frac{\Omega_{0,s}}{|\langle \mathbf{B} \rangle|}\right)^2 V_{\rm sw}\left(r\right) P_{{\rm B},xx}\left(f, r\right).
    \label{eq:D_mumu}
\eeq
Here, $v$ is the particle speed, $\mu = v_\parallel/v$ is the cosine of the pitch angle, $\Omega_{0,s}$ is the gyrofrequency of GCRs of the species $s$, $f \approx {V_{\rm sw} \Omega_{0,s}}/\left( 2 \pi \mu v \right)$ is the frequency of the hydromagnetic waves that GCRs of the species $s$ at pitch-angle cosine $\mu$ resonantly interact with, and the subscript ``$x$'' of $P_{{\rm B}, {xx}}\left(f, r\right)$ is one of the directions normal to the mean magnetic field $\langle \mathbf{B} \rangle$. Because the \emph{N} direction at the solar ecliptic plane is perpendicular to $\langle \mathbf{B} \rangle$, we have $P_{{\rm B}, {xx}}\left(f, r\right) = P_{{\rm B}, {\rm NN}}\left(f, r\right)$. In Equation~\eqref{eq:diffusion_QLT}, $\mu_{{\rm min},s}$ is set at where the resonant frequency of the species $s$, ${V_{\rm sw} \Omega_{0,s}}/ \left( 2 \pi \mu v \right)$, equals $f_{\rho p}$. In Equation~\eqref{eq:D_mumu}, Taylor's hypothesis~\citep{1938RSPSA.164..476T} has been used to convert the wavenumber $k$ to $f$ through $f = k V_{\rm sw} / 2 \pi$.

The assumption of using Taylor's hypothesis near the Sun is expected to remain a good approximation if the sampling angle, defined as the angle between the spacecraft's motion and the the local magnetic field, is greater than $30^\circ$~\citep{2021A&A...650A..22P}. At 0.1--0.3~au, PSP moves nearly perpendicular to the local magnetic field, and thus we can use Taylor's hypothesis to reconstruct the spatial energy spectra from the measured frequency spectra. Far away from the Sun where the solar wind speed is much larger than the rms speed of the fluid and the \Alfven speed, Taylor's hypothesis is valid regardless of the sampling angle.

\begin{figure*}[t] 
   \centering
   \includegraphics[width=0.48\textwidth]{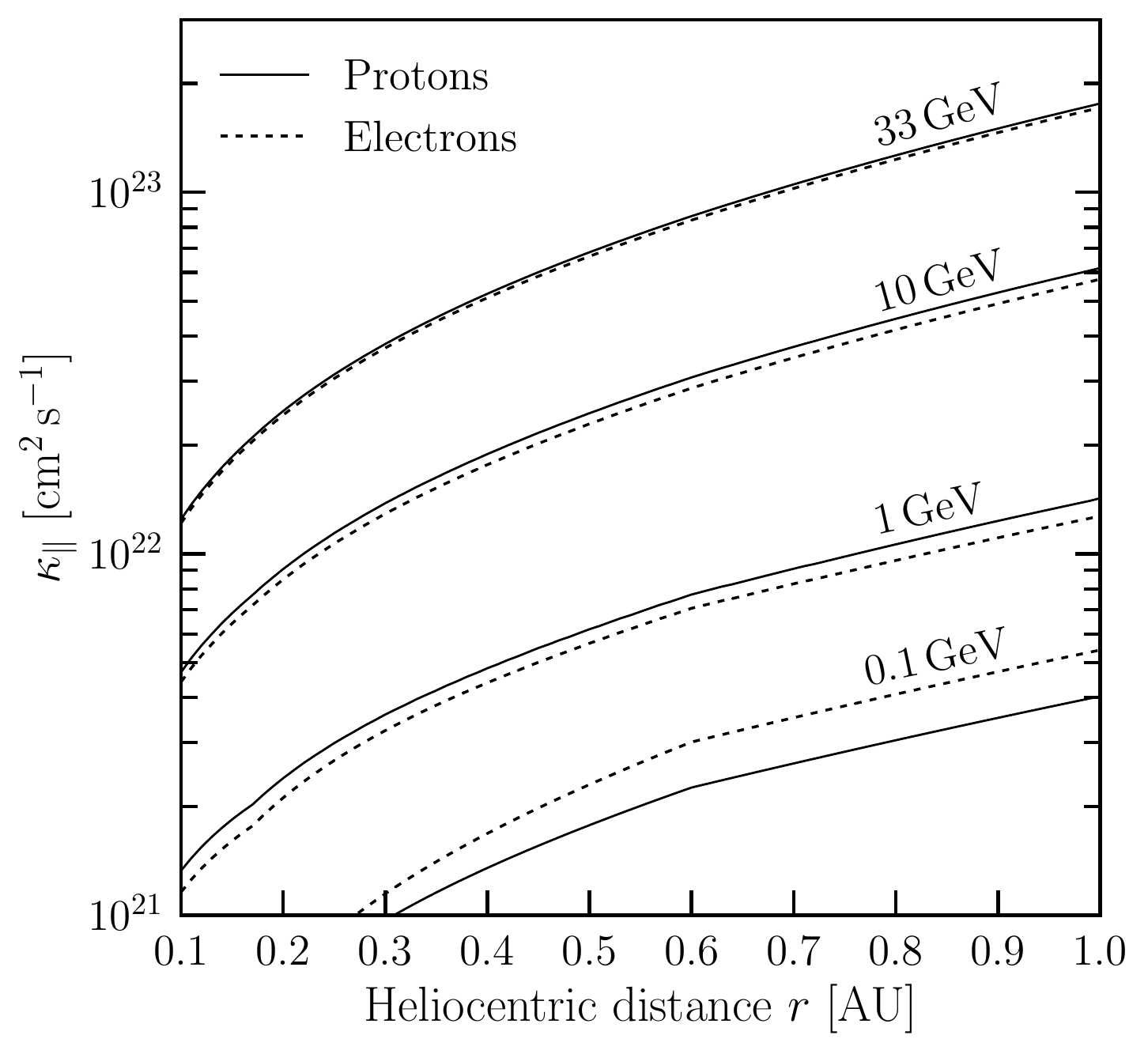} \includegraphics[width=0.48\textwidth]{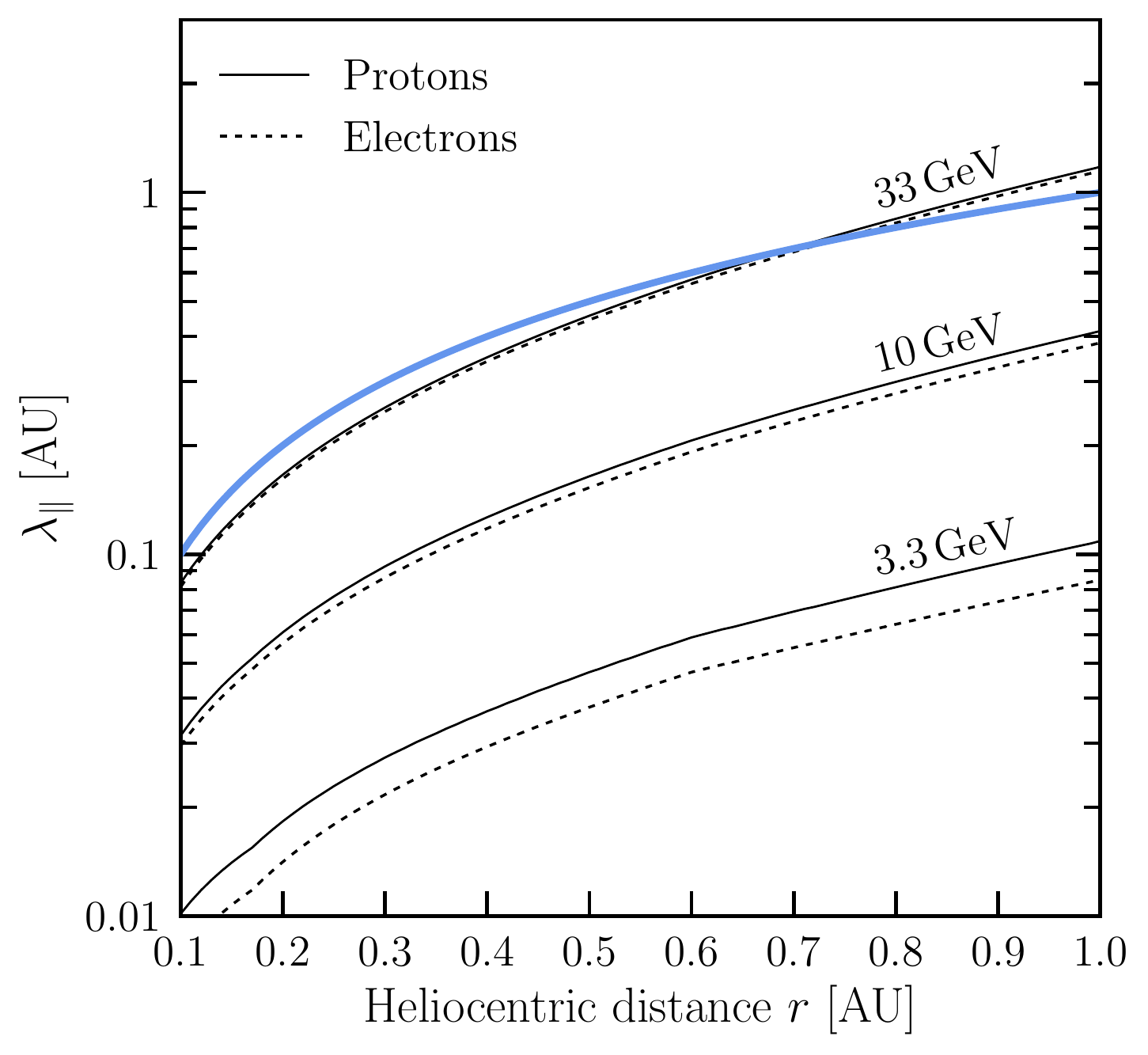} 	
   \caption{Parallel diffusion coefficient (left), $\kappa_\parallel$, and parallel mean free path (right), $\lambda_\parallel$, as a function of heliocentric distance $r$ for various particle kinetic energies, $T$. The blue line in the right plot is where $\lambda_\parallel = r$; above this line (approximately for $T \gtrsim 33 \: {\rm GeV}$), the diffusion description of the cosmic-ray transport in the inner heliosphere is not strictly valid.}
   \label{fig:kappa_r}
\end{figure*}

\begin{figure*}[tpb] 
   \centering
   \includegraphics[width=0.48\textwidth]{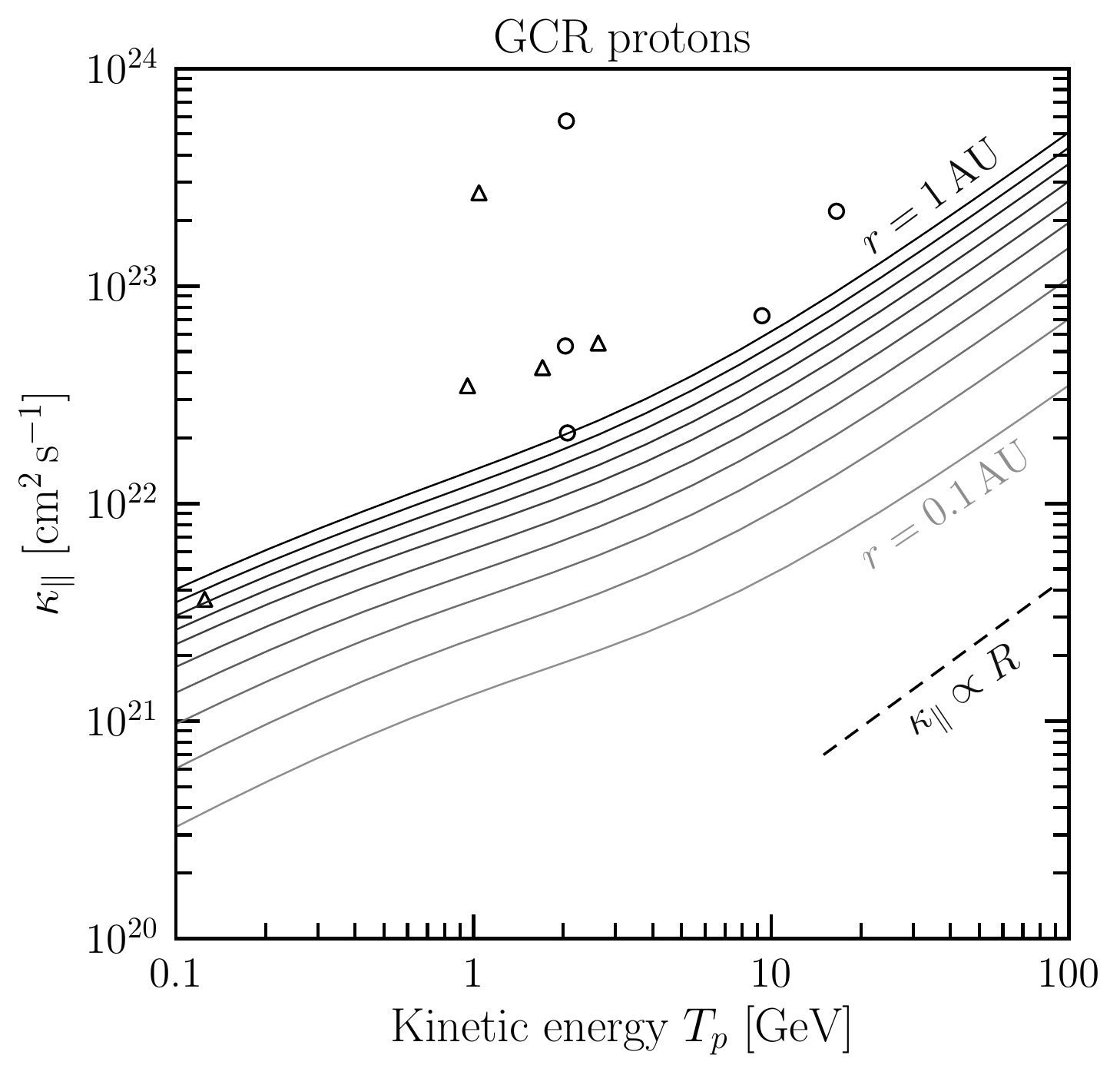}
   \includegraphics[width=0.48\textwidth]{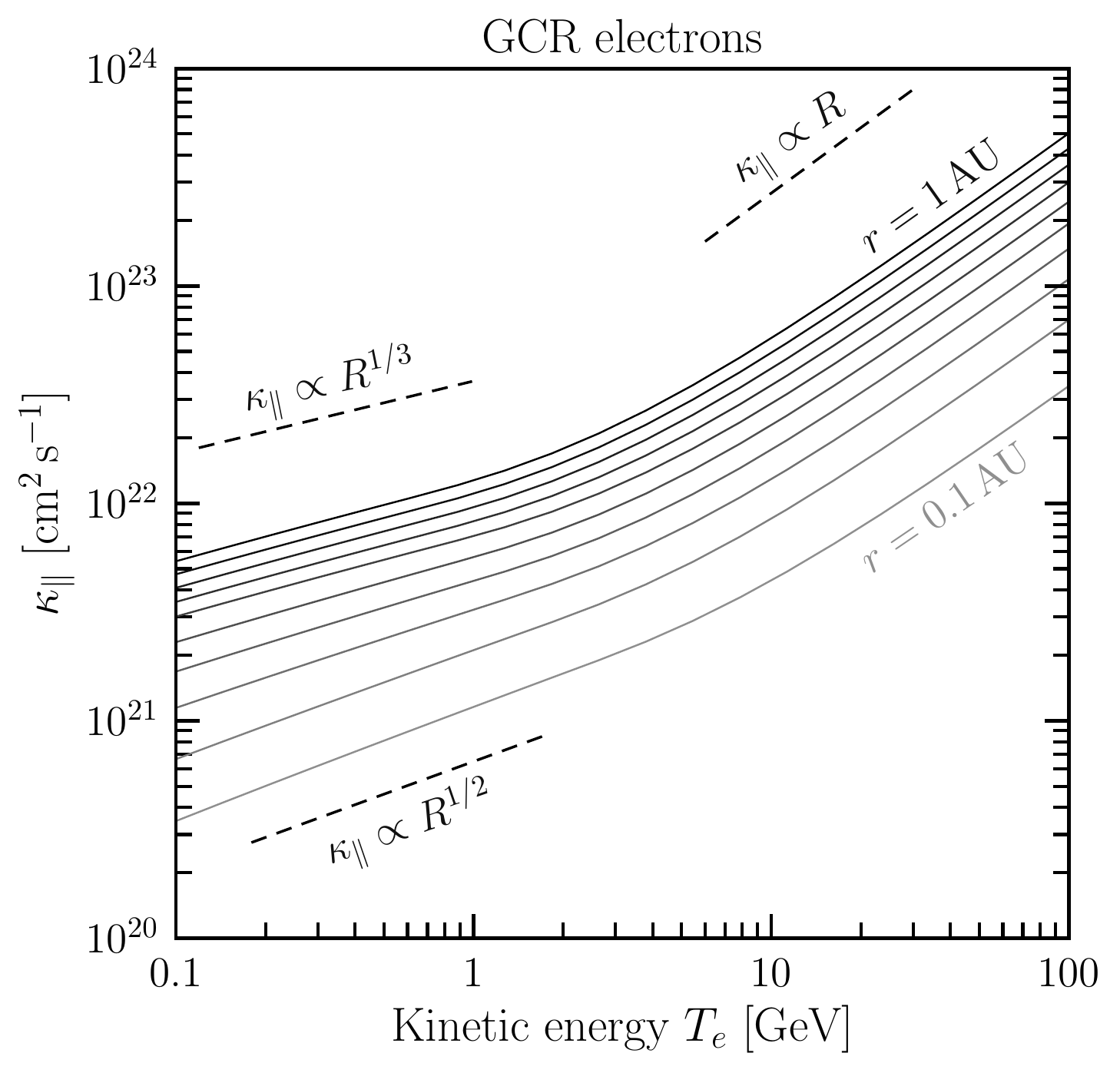}	
   \caption{Parallel diffusion coefficient of protons (left) and electrons (right) as a function of kinetic energy $T$ for heliocentric distances between 0.1 and 1~au, spaced in intervals of 0.1~au. The open circles and upward-pointing triangles are derived from the actual and lower-limit values of the measurements of $\lambda_\parallel$ at the heliocentric distance of 1~au, respectively.}
   \label{fig:kappa_T}
\end{figure*}

\begin{figure*}[t] 
   \centering
   \includegraphics[width=0.48\textwidth]{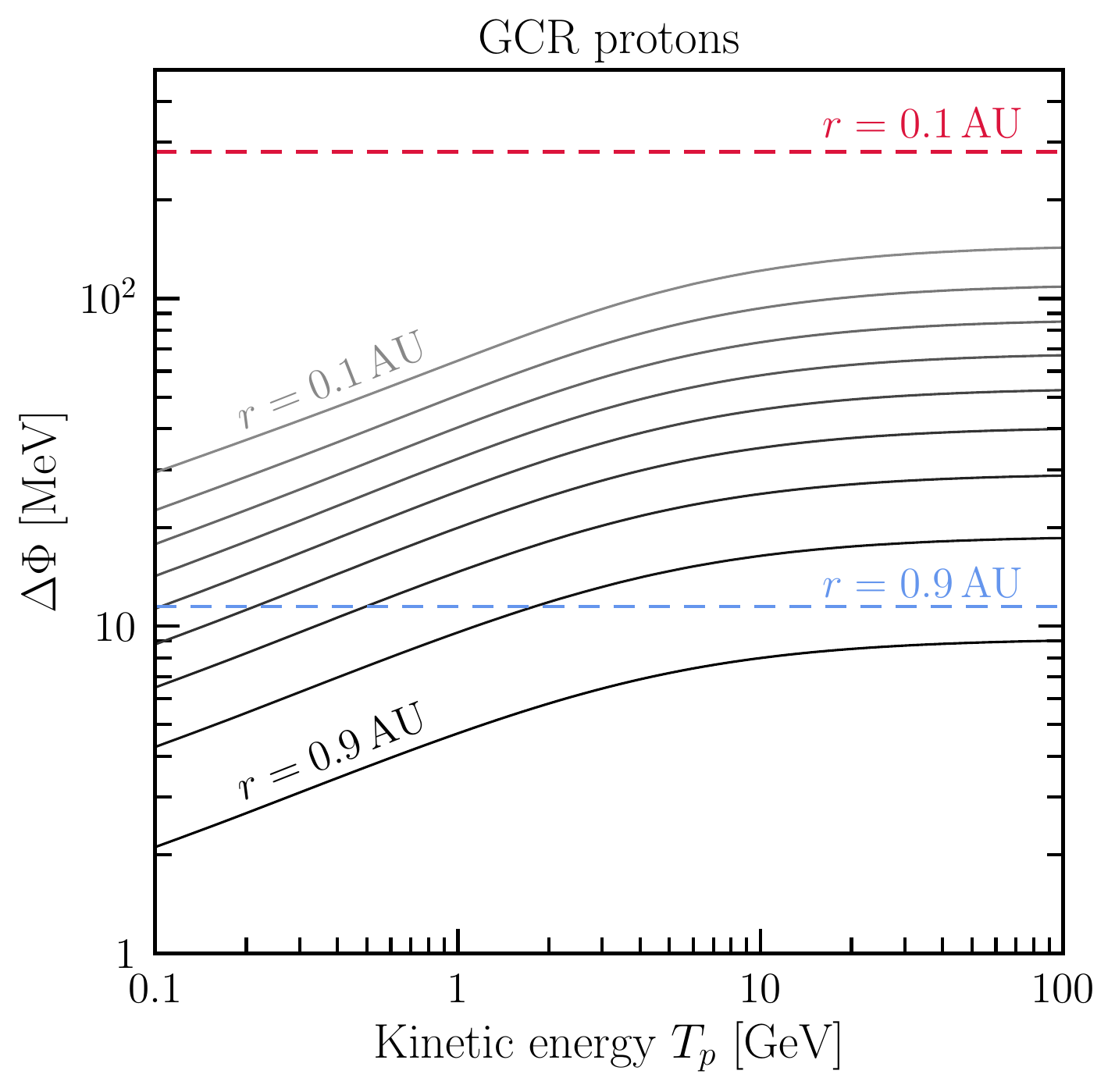}
   \includegraphics[width=0.48\textwidth]{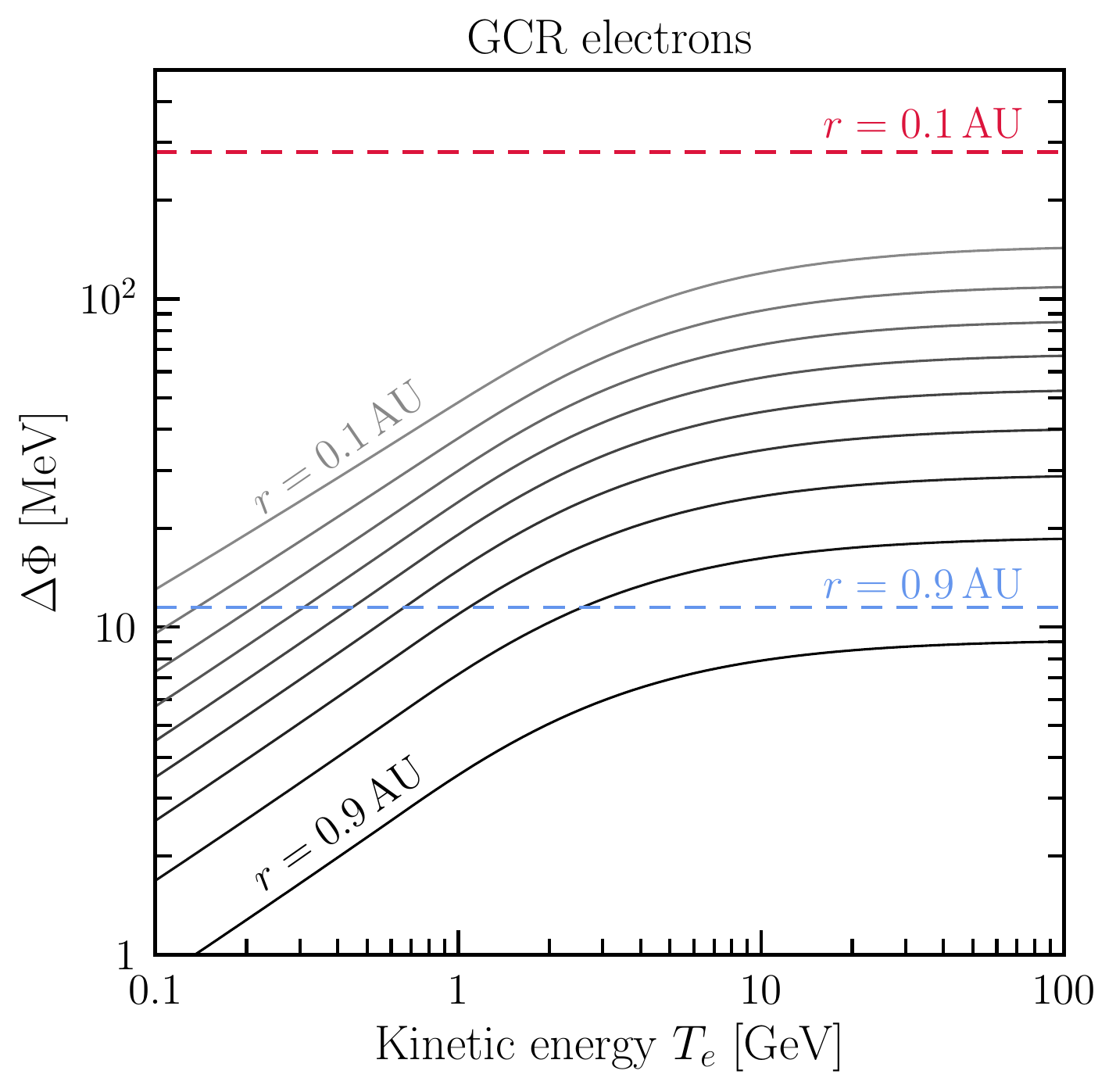}
   \caption{Calculated growth of the rigidity-dependent $\Phi$ (solid lines) from $1\,{\rm au}$ to radii $r$ between 0.1 and 0.9~au in steps of 0.1~au. As a comparison, we show the growth of the rigidity-independent $\Phi$ for $r=0.1\,{\rm au}$ (top dashed line) and $r=0.9\,{\rm au}$ (bottom dashed line).}
   \label{fig:modulation_potential}
\end{figure*}

To determine $\kappa_\parallel$ at different locations in the heliosphere, we need to know the profiles of $\langle \mathbf{B} \rangle$ and $V_{\rm sw}$ as functions of $r$. First, we adopt Parker's Archimedean spiral magnetic field model for $\langle \mathbf{B} \rangle$ (i.e., the IMF), which gives~\citep{1958ApJ...128..664P},
\beq
    \langle \mathbf{B} \rangle = A B_r\left(r\right) \left[ \hat{\mathbf{e}}_r - \hat{\mathbf{e}}_\phi \frac{\left(r-r_\odot\right) \Omega_\odot \sin\theta}{V_{\rm sw}\left(r\right)} \right].
    \label{eq:IMF}
\eeq
Here, $A$ is the sign of the solar polarity, $r_\odot \approx 6.96\times 10^{10}~{\rm cm}$ is the solar radius (throughout the text, we use lowercase $r_\odot$ to denote the solar radius), $B_r\left(r\right) = 2.83 \, {\rm nT} \times \left({r}/{\rm au}\right)^{-2}$ is the radial component of the Archimedean magnetic field at solar minimum, $\Omega_\odot \approx 2.97 \times 10^{-6}~{\rm rad~s^{-1}}$ is the solar rotation speed, and $\left(r, \theta, \phi \right)$ are the spherical coordinates relative to Sun's rotation axis. Our choice of $B_r\left(1\:{\rm au}\right)=2.83~{\rm nT}$ is taken from the mean radial magnetic field strength measured by Ulysses' first full polar orbit during the low solar activity period~\citep{2000JGR...10510419M}. This choice of $B_r$ leads to the total magnetic field at 1~au at the solar ecliptic being $\approx 4.23~{\rm nT}$, which agrees with the measurements at solar minimum in the past few solar cycles~\citep{1993AdSpR..13f..15B, 2015PEPS....2...13G, 2017SSRv..212.1271K}.

With the IMF given in Equation~\eqref{eq:IMF}, we write the Parker spiral angle $\psi$ in the solar ecliptic plane as
\beq
    \psi\left(r\right) = \tan^{-1} \bigg[ \frac{\left(r-r_\odot\right) \Omega_\odot}{V_{\rm sw}\left(r\right)} \bigg] .
\eeq

For the solar wind speed, we adopt the empirical model from \cite{2006SSRv..127..117H}, which expresses $V_{{\rm sw}}$ along the solar ecliptic plane as
\beq
    V_{{\rm sw}}\left(r\right) = V_0 \: \bigg[ 1 - \exp \left( \frac{40}{3} \left( \frac{r_\odot -r}{\rm 1\: au} \right) \right) \bigg],
\eeq
with $V_0 = {\rm 400~km/s}$. This model agrees well with the observations from SOHO showing the wind in the ecliptic plane typically accelerates from the rest to $300~{\rm km/s}$ at $25~r_\odot$ to its maximum speed of $400~{\rm km/s}$ at $0.3~{\rm au}$, after which the speed remains nearly constant~\citep{1997ApJ...484..472S}.

\subsection{Parallel Diffusion Results}\label{subsec:parallel_diffusion_result}

Figure~\ref{fig:kappa_r} shows our numerical results of $\kappa_\parallel$ (top left) and particle mean free path $\lambda_\parallel = 3 \kappa_\parallel / v$ (top right) as a function of $r$. The solid lines are for GCR protons and the dashed lines are for to GCR electrons. The blue line in the right plot is where $\lambda_\parallel$ equals the characteristic size of the system, $r$. For $T\gtrsim 33~{\rm GeV}$, we have $\lambda_\parallel > r$, so the normal diffusion approximation of the particle transport is not valid. Combining this point with the constraint due to the lack of accurate PSD measurements below $2\times 10^{-6}~{\rm Hz}$, we consider our analysis as strictly valid for $T \lesssim 33~{\rm GeV}$, which is adequate because modulation at higher energies is small.

In Figure~\ref{fig:kappa_r}, it is interesting that protons and electrons have essentially the same $\kappa_\parallel$ for $T\gtrsim 10~{\rm GeV}$. This is because the resonant frequency, $f \approx {V_{\rm sw} \Omega_{0,i}}/{ 2 \pi \mu v}$, is linearly proportional to $R/v$ and thus becomes independent of the rest masses of the proton and electron, $m_p$ and $m_e$, whenever $T\gg m_p c^2, \, m_e c^2$. At low energies, however, protons and electrons with the same $T$ have different $\kappa_\parallel$. In particular, we see that at $T=0.1~{\rm GeV}$, $\kappa_\parallel$ for a relativistic electron is larger than $\kappa_\parallel$ for a nonrelativistic proton. The reason is twofold: first, $\kappa_\parallel$ is an increasing function of $v$. Second, the $R$ of electrons is lower than the $R$ of protons provided they have the same $T$; this indicates that an electron resonantly scatters a slightly smaller PSD than a proton does.

Figure~\ref{fig:kappa_T} shows $\kappa_\parallel$ for GCR protons (bottom left) and electrons (bottom right) as a function of $T$. The low-energy ($\lesssim 1 \: {\rm GeV}$) and the high-energy ($\gtrsim 10 \: {\rm GeV}$) regimes have different slopes because GCRs at these two different energy regimes resonantly interact with different types of magnetic turbulence. QLT suggests that $\kappa_\parallel$ for the relativistic particles in Equation~\eqref{eq:diffusion_QLT} should vary as $R^{2+\nu}$ for a magnetic power spectrum that varies as $k^{\nu}$. For a Kolmogorov-like, $P_{xx}\propto k^{-5/3}$, and an Iroshnikov--Kraichnan-like, $P_{xx}\propto k^{-3/2}$~\citep{1964SvA.....7..566I, 1965PhFl....8.1385K}, \Alfven wave turbulence, $\kappa_\parallel$ varies as $R^{1/3}$ and $R^{1/2}$, respectively, as shown in the low-energy regime. For an $1/f$ fluctuation ($P_{xx}\propto k^{-1}$), $\kappa_\parallel$ varies as $R$, as shown in the high-energy regime.

In the left plot of Figure~\ref{fig:kappa_T}, the data points are obtained from the measurements of $\lambda_\parallel$ at $r=1~{\rm au}$ summarized in~\cite{1982RvGSP..20..335P} and~\cite{1994ApJ...420..294B}. Specifically, these data are reported in~\cite{1977SoPh...54..207S}, \cite{1977JGR....82.4704F}, \cite{1978JGR....83.1157Z}, \cite{1983GeoRL..10..920B}, \cite{1986JGR....91.8713B}, \cite{1987ApJ...322.1052B}, and~\cite{1993ApJ...405..375C}. We see that for $1~{\rm GeV} < T_p < 20~{\rm GeV}$, the theory prediction at $r=1$~au is at least a factor $2$ smaller than the data points. While we do not show it here, the discrepancy is even larger at $T_p \lesssim 0.1~{\rm GeV}$ where the theoretical prediction can be smaller than the data points by a factor of $\simeq 10$. The measurements of electron $\lambda_\parallel$ also show similar discrepancies at $T_e\lesssim 10~{\rm GeV}$. This issue, known as the Palmer consensus, is reported in~\cite{1982RvGSP..20..335P}. The Palmer consensus indicates that the true cosmic-ray interaction with magnetic turbulence in the inner heliosphere is \emph{weaker} than the predictions from the standard QLT in a magnetostatic, dissipationless turbulence with slab geometry~\citep{1966ApJ...146..480J}. 

Throughout this paper, we still use the standard QLT treatment in Equations~\eqref{eq:diffusion_QLT} and~\eqref{eq:D_mumu} to derive $\kappa_\parallel$ unless otherwise specified. We note that because realistic cosmic-ray scattering with magnetic turbulence is weaker than the QLT-predicted values, the true modulation is likely smaller than that derived in this work. For example, the measured $\kappa_\parallel$ of proton at $1~{\rm GeV} < T_p < 20~{\rm GeV}$ is about twice the QLT-predicted value, as shown in Figure~\ref{fig:kappa_T}. According to the characteristic equation in Equation~\eqref{eq:characteristic_eqn}, the change in $\Phi$ in a realistic solar wind environment would be roughly half of the change in $\Phi$ calculated from the standard QLT prediction.

\subsection{Modulation Potential Energy}\label{subsec:modulation_potential}

We calculate the growth of the modulation potential energy, $\Delta \Phi \equiv \Phi\left(r\right) - \Phi\left(1~{\rm au}\right)$, from $1\:{\rm au}$ to inner heliospheric radii $r$ by applying the numerical results of $\kappa_\parallel$ in the force-field characteristic equation in Equation~\eqref{eq:characteristic_eqn}. We note that $\Delta \Phi$ is positive because smaller $r$ has larger $\Phi$.

Figure~\ref{fig:modulation_potential} shows the calculated $\Delta \Phi$ (solid lines) for GCR protons and GCR electrons. At $T \lesssim 10\:{\rm GeV}$, $\Delta \Phi$ for each $r$ is an increasing function of $T$, and GCRs at this energy range have small enough gyro-radii that they pitch-angle scatter with the inertial-range turbulence. At $10\:{\rm GeV} \lesssim T \lesssim 33\:{\rm GeV}$, $\Delta \Phi$ for each $r$ reaches a plateau and becomes rigidity (or energy) independent because GCRs in this energy range predominantly pitch-angle scatter with the $1/f$ fluctuations. Since the magnetic power in the $1/f$ range is higher than the power in the inertial range, scattering with $1/f$ fluctuations results in a stronger solar modulation, as evidenced by the trend of the solid lines. We note again that at $T\gtrsim 33\:{\rm GeV}$, our analysis does not hold because (i) the spectral shape of the magnetic PSD below $f\lesssim 2\times 10^{-6}\,{\rm Hz}$ is not known and (ii) the normal diffusion approximation does not apply.

In Figure~\ref{fig:modulation_potential}, we also show $\Delta \Phi$ from the rigidity-independent $\Phi$ model in Equation~\eqref{eq:R_indep_compact} assuming $\earthpotential = 400 \, {\rm MeV}$, a constant $V_{\rm sw}$, and $\eta = 1.1$. (Note that our solar wind model has radial dependence, which has non-negligible effects on $\Delta \Phi$ for $r \lesssim {\rm 0.3 \, au}$.) Our choice of $\earthpotential = 400 \, {\rm MeV}$ during solar minimum is consistent with the neutron monitor measurements reported in~\cite{2017JGRA..122.3875U}. At $T \gtrsim 10 \, {\rm GeV}$, our model and the rigidity-independent $\Phi$ model differ by a factor of $2$. At $T \lesssim 10 \, {\rm GeV}$, the gap between the two models is significant. In particular, $\Delta\Phi$ in the rigidity-independent case is overestimated by one order of magnitude at $T \sim 0.1 \, {\rm GeV}$.

Last, we remark that the rigidity dependence of $\Phi$ in this work is different from that in \cite{2016PhRvD..93d3016C}, \cite{2016ApJ...829....8C}, and \cite{2017JGRA..12210964G}. Here, we predict the increase of $\Phi$ from 1~au to inner heliocentric radii, whereas they fit the direct GCR measurements at 1~au with the local interstellar spectrum of GCRs using the rigidity-dependent parameterization of $\earthpotential$. In particular, the rigidity dependence of $\Phi$ in our model stems entirely from GCR interactions with the inertial-range turbulence. On the other hand, their rigidity-dependent parameterization of $\earthpotential$ is a mixture of all possible sources of solar modulation, from the heliospheric boundary down to 1~au.

\section{Predictions}\label{sec:result}

In this section, we show our results for the GCR intensities and radial gradients in the inner heliosphere. In Section~\ref{subsec:radial_gradient}, we compare our calculated GCR radial gradients with the measurements from Helios and Pioneer missions in the solar ecliptic plane within 5~au from the Sun. This is a check on our model vs. the rigidity-independent $\Phi$ model. In Section~\ref{subsec:GCR_flux}, we show the GCR proton and electron energy spectra inside 1~au.

\subsection{Radial Gradients}\label{subsec:radial_gradient}

To check the validity and applicability of our improved force-field model, we calculate the radial gradients $G_r$ and compare them with the $G_r$ measurements of GCR protons at $0.3 \,{\rm au} < r < 1\,{\rm au}$~\citep{2019A&A...625A.153M} and GCR helium nuclei at $1 \,{\rm au} < r < 4.6\,{\rm au}$~\citep{1977ApJ...216..930M}.

The radial gradient $G_r$ is defined as
\beq
    G_r \equiv \frac{\ln I\left(r_2\right)- \ln I\left(r_1\right)}{r_2 - r_1},
    \label{eq:Gr_formula}
\eeq
where $I\left(r\right) = \int_{T_1}^{T_2} J_E\left(T, r\right) dT$ is the integrated GCR intensity over a range of kinetic energy $[T_1, T_2]$ at the heliocentric distance $r$. To obtain $J_E$ at $r<{\rm 1~au}$, we use the force-field solution from Equation~\eqref{eq:force_field_solution},
\beq
    {J_E\left(E, r\right)} = \left( \frac{E^2-E_0^2}{\left(E+\Delta\Phi\right)^2 - E_0^2} \right) {J_E\left(E+\Delta\Phi, {\rm 1\,au}\right)},
    \label{eq:GCR_intensity_02AU}
\eeq
where $J_E\left(E+\Delta\Phi, {\rm 1~au}\right)$ is the GCR intensity at 1~au.

In this work, we use the PAMELA observations of protons, electrons, and helium nuclei in 2009-2010 for $J_E$ at 1~au~\citep{2013ApJ...765...91A, 2015ApJ...810..142A, 2020ApJ...893..145M}. We note more recent AMS observations of proton and helium nuclei in 2018-2019 reported in~\cite{2021PhRvL.127A1102A, 2022PhRvL.128w1102A}. Here, we did not use the AMS results because the lowest particle kinetic energy provided in the AMS observations is above the particle kinetic energies in the radial-gradient measurements reported in the Helios and Pioneer missions. We emphasize that the choice of different GCR spectra data at 1~au is less important to the results shown in this work as our focus is on the relative change in GCR spectra from Earth.

\begin{table}[t]
\center
\caption{Radial gradients of GCR protons (top group) and helium nuclei (bottom group). The second column, $T_{\rm nucleon}$, denotes the kinetic energy per nucleon. In the last column, the numerical values without and inside the parentheses are based on the standard QLT treatment of $\kappa_\parallel$ and the modification of $\kappa_\parallel$, respectively.}
\begin{tabular}{cccc}
\toprule%
$r$ & $T_{\rm nucleon}$ & Measured $G_r$ & Calculated $G_r$ \\ 
$\left({\rm au}\right)$ & $\left({\rm GeV}\right)$ & $\left(\% \: {\rm au^{-1}}\right)$ & $\left(\% \: {\rm au^{-1}}\right)$ \\ 
\toprule%
0.3--1 & 0.250--0.700 & $2 \pm 2.5$ & 14.6 (7.4) \\ 
0.4--1 & $>$0.05 & $6.6 \pm 4$ & 12.6 (6.2)\\
\midrule%
1--3.8 & 0.210--0.275 & $0 \pm 4$ & 9.2 (4.6) \\ 
1--3.8 & 0.275--0.380 & $2.5 \pm 4$ & 10.1 (4.9) \\ 
1--3.8 & 0.380--0.460 & $3.8 \pm 5$ & 10.9 (5.3) \\ 
1.25--4.2 & 0.210--0.275 & $4.1 \pm 3.7$ & 9.1 (4.6) \\ 
1.25--4.2 & 0.275--0.380 & $2 \pm 4$ & 10.1 (4.8) \\ 
1.25--4.2 & 0.380--0.460 & $1.3 \pm 5$ & 10.8 (5.3) \\ 
1.9--4.6 & 0.210--0.275 & $2.7 \pm 4$ & 8.7 (4.5) \\
1.9--4.6 & 0.275--0.380 & $2.5 \pm 5$ & 10.0 (4.7) \\
1.9--4.6 & 0.380--0.460 & $0 \pm 5$ & 10.7 (5.3) \\
\bottomrule%
\end{tabular}
\label{table:radial_gradient}
\end{table}

\begin{figure}[t] 
   \centering
   \includegraphics[width=0.48\textwidth]{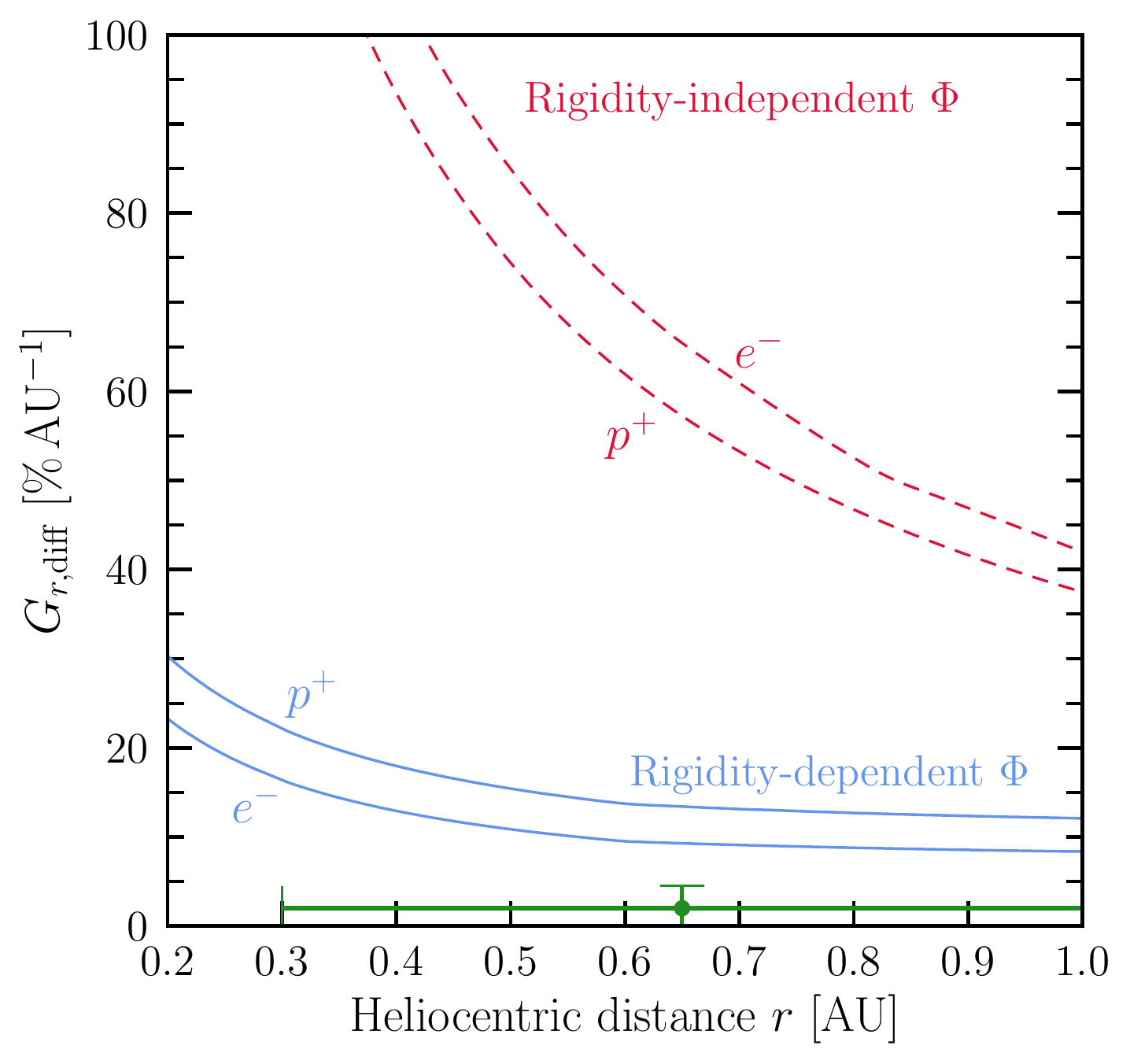} 
   \caption{Calculated differential radial gradients $G_{r,{\rm diff}}$ of GCR protons and electrons for $T=[0.25, 0.7]~{\rm GeV}$. Blue solid lines indicate our results for the standard QLT treatment. Red dashed lines indicate the rigidity-independent $\Phi$ model. The green datum is the measured $G_r$ of GCR protons of $2\% \pm 2.5\%~{\rm au^{-1}}$ between 0.3~au and 1~au obtained from the Helios Experiment-6~\citep{2019A&A...625A.153M}.}
   \label{fig:gradient}
\end{figure}

\begin{figure*}[t]
   \centering
   \includegraphics[width=0.48\textwidth]{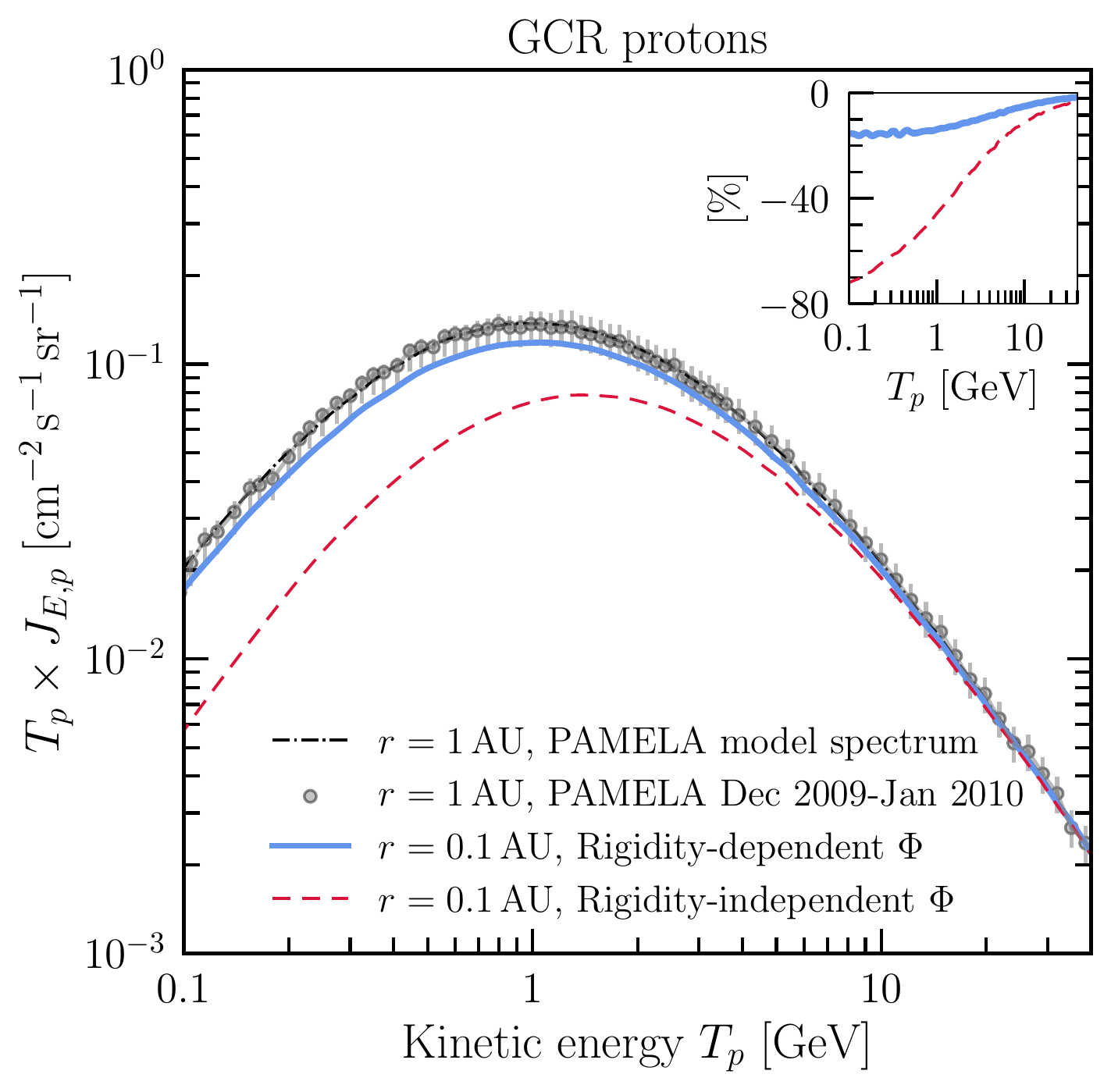}
   \includegraphics[width=0.48\textwidth]{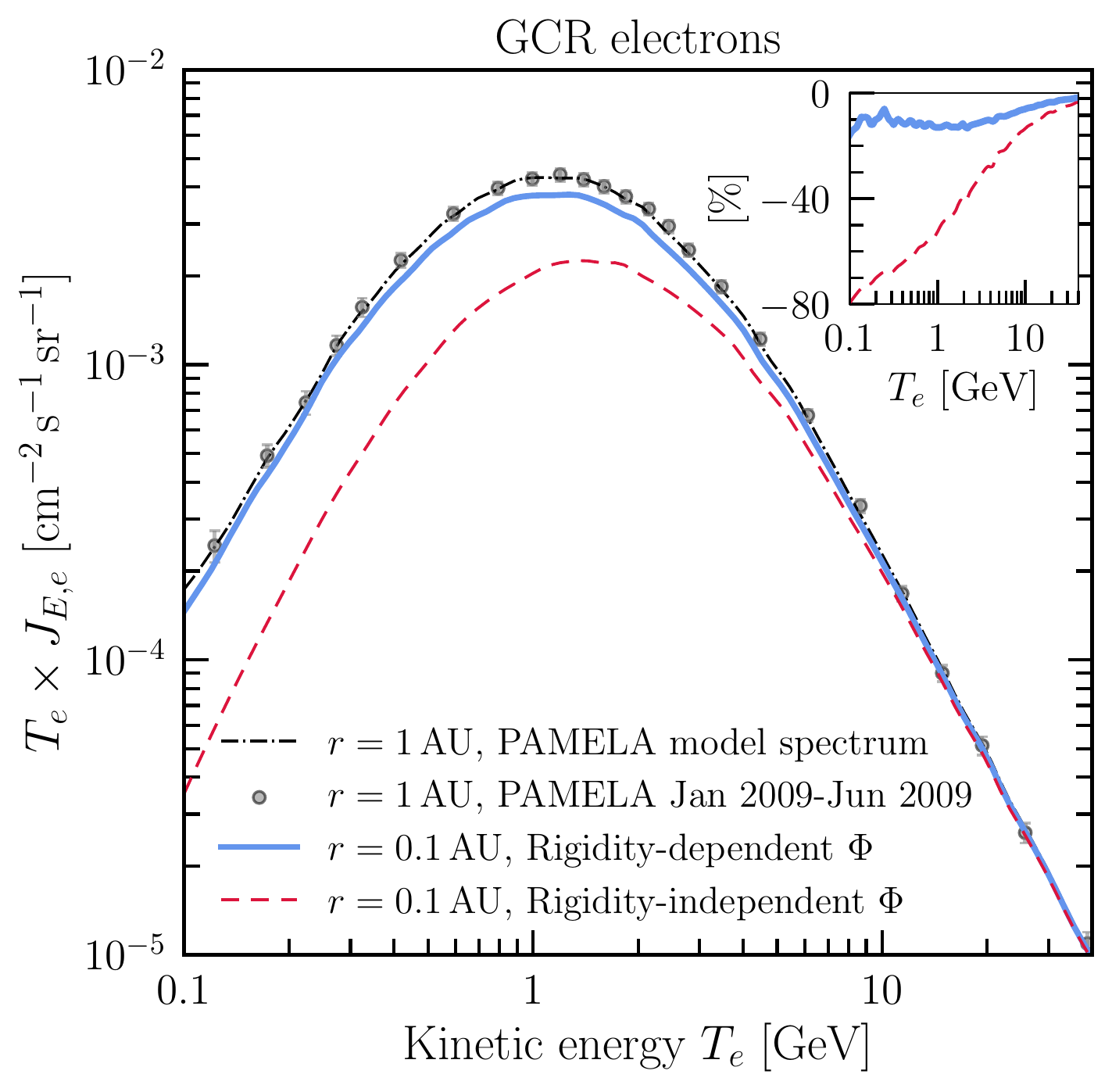}
   \caption{Predicted spectra of GCR protons (left) and electrons (right) at $r=0.1\:{\rm au}$. The blue solid lines indicate our results. The red dashed lines indicate the rigidity-independent $\Phi$ model. The gray circles are the PAMELA proton and electron observations with the error bars denoting the combined statistical and systematic errors, and the black dash--dotted lines showing their model spectra~\citep{2013ApJ...765...91A, 2015ApJ...810..142A}. Inset: the percentage change of GCR intensity from 1--0.1~au.} 
   \label{fig:CR_flux}
\end{figure*}

Table~\ref{table:radial_gradient} lists the measured $G_r$ of proton (top group) from Helios~1 and~2 missions during 1974--1978~\citep{2019A&A...625A.153M} and $G_r$ of helium nuclei (bottom group) from Pioneer~10, Pioneer~11, and Helios~1 during 1973--1975~\citep{1977ApJ...216..930M}. Both groups were observed during the solar minimum at the end of the Solar Cycle~20. In the last column for the calculated $G_r$, the numerical values without the parenthesis are based on the standard QLT treatment for $\kappa_\parallel$ in Equation~\eqref{eq:diffusion_QLT}. Comparing them with the measured $G_r$, we find that our results for the GCR protons for $r<1$~au are $\simeq$~$6\%$--$12\%$~${\rm au^{-1}}$ away from the measurements. Our results for the GCR helium nuclei at ${\rm 1~au} < r < {\rm 4.6~au}$ are $\simeq$~$5\%$--$8\%$~${\rm au^{-1}}$ away from the measurements.

In Table~\ref{table:radial_gradient}, the numerical values of $G_r$ inside the parenthesis in the last column are based on doubling the standard QLT treatment of $\kappa_\parallel$ in Equation~\eqref{eq:diffusion_QLT}. This choice is motivated by the observations in~\cite{1982RvGSP..20..335P} showing that the observed proton mean free paths for $1~{\rm GeV} < T_p < 20~{\rm GeV}$ are approximately twice the predicted values from the standard QLT, as discussed in Section~\ref{subsec:parallel_diffusion_result}. Based on this modification of $\kappa_\parallel$, we find that the calculated $G_r$ for GCR proton and helium nuclei agree well with the measurements.

In Figure~\ref{fig:gradient}, we show our calculations of differential radial gradients, $G_{r,{\rm diff}} \equiv {\partial \ln I\left(r\right)}/{\partial r}$, for GCR protons and electrons in the kinetic energy range $[0.25, 0.7]$~GeV. This set of calculations is based on the standard QLT treatment of $\kappa_\parallel$ in Equation~\eqref{eq:diffusion_QLT}. For comparison, we also show $G_{r,{\rm diff}}$ using the rigidity-independent $\Phi$ model in Equation~\eqref{eq:R_indep_compact}, assuming $\earthpotential=400\:{\rm MeV}$ and $\eta={1.1}$. Within 1~au, we find that $G_{r,{\rm diff}}$ of the rigidity-independent $\Phi$ model is higher than the $G_{r,{\rm diff}}$ of our model by at least a factor of 5. It is apparent that rigidity independence of $\Phi$ leads to an over-modulation of the low-energy GCR, especially as particles approach the Sun.

Last, our analysis of $G_r$ has combined measurements from three different solar cycles: $G_r$ from Cycle 20, the GCR spectra at 1~au from Cycle 24, and the magnetic PSD from Cycle 25. This could potentially lead to an inconsistency between the calculated and the measured $G_r$. In particular, $G_r$ is highly sensitive to the magnetic PSD and frequency break. The magnetic condition of the solar wind may be very different at the time $G_r$ was measured (1974--1978) and the time PSP took data (2018--2019). A more consistent analysis can be obtained in the near future once PSP and SolO release the measurements of $G_r$ and magnetic PSD from the same solar cycle.

\subsection{Galactic Cosmic-Ray Intensity}\label{subsec:GCR_flux}

In this subsection, we calculate $J_E$ for GCR protons and electrons at $r<1~{\rm au}$ from the force-field solution in Equation~\eqref{eq:GCR_intensity_02AU}. Throughout the calculation, we have adopted the numerical values of $\Delta \Phi$ from Figure~\ref{fig:modulation_potential}. We also used the dashed--dotted model spectral lines in Figure~\ref{fig:CR_flux} as the input for $J_E\left(E+\Delta\Phi, {\rm 1~au}\right)$. Our calculation of $J_E$ of GCR helium nuclei is presented in the Appendix. (We note that the calculations shown in this subsection and in the Appendix are based on the standard QLT treatment of $\kappa_\parallel$ in Equation~\eqref{eq:diffusion_QLT}.)

Figure~\ref{fig:CR_flux} shows our results (blue solid lines) for the GCR proton and electron energy spectra at $r=0.1~{\rm au}$. We see that the solar modulation accumulated from $1~{\rm au}$ down to $0.1\:{\rm au}$ leads to no more than $\simeq -16\%$ of GCR intensity reduction at $0.1~{\rm GeV} \lesssim T \lesssim 10~{\rm GeV}$ and no more than $\simeq -5\%$ intensity reduction at $10~{\rm GeV} \lesssim T \lesssim 40~{\rm GeV}$. While the GCR spectra for $0.1\,{\rm au} < r < 1\,{\rm au}$ are not shown here, we have confirmed that the spectra lie between the black and blue lines.

In Figure~\ref{fig:CR_flux}, we also show in red dashed lines the GCR energy spectra from the rigidity-independent $\Phi$ model in Equation~\eqref{eq:R_indep_compact}, assuming $\earthpotential = 400~{\rm MeV}$ and $\eta={1.1}$. We see that the red dashed lines reach $\simeq -70\%$ to $-80\%$ at $T=0.1~{\rm GeV}$ as compared to the blue lines of $\simeq -10\%$ to $-15\%$ at the same $T$. The red dashed lines eventually increase to the same level as the blue solid lines by $T=40~{\rm GeV}$, at which energy the solar modulation is already negligible.

In summary, our calculations show that the modulation in the inner heliosphere is modest. We demonstrate that the GCR spectra down to 0.1~au are close to those measured at 1~au. Our results also suggest that the gamma-ray emission from the solar halo due to electron GCR scattering with solar photons should be higher than that of previous predictions.

\section{Discussion and Conclusions}\label{sec:conclusion}

An important goal of heliophysics is to understand the propagation of charged cosmic rays toward the Sun in the magnetically turbulent solar wind. In this paper, we have presented an improved force-field model for calculating the modulation potential energy and GCR intensity in the inner heliosphere whenever the magnetic PSD in the solar wind is known. The magnetic PSD adopted in this study reflects the solar wind conditions in the solar ecliptic plane during the solar minimum at the end of Solar Cycle 24.

We show that the increase of the modulation potential energy, $\Delta \Phi$, from 1~au to inner heliocentric radii is rigidity dependent at kinetic energies $T\lesssim 10~{\rm GeV}$ due to GCR interactions with inertial-range turbulence. At $10~{\rm GeV} \lesssim T \lesssim 33~{\rm GeV}$, $\Delta \Phi$ is independent of rigidity due to GCR interactions with $1/f$ fluctuations. Overall, we find a modest reduction of GCR intensity at low particle kinetic energies in the inner heliosphere.

The same method can be applied to the GCR modulation in the solar ecliptic plane at solar maximum. However, the magnetic spectral shape of the solar wind at solar maximum could be quite different from that of solar minimum. This is because the solar wind at solar maximum is mostly slow-wind streams whereas the wind at solar minimum is mixed, with fast- and slow-wind streams~\citep{1995SSRv...73....1T}. A slow-wind stream has the frequency break occurring at much lower frequencies than a fast wind or a mixed configuration~\citep{2009EM&P..104..101B}. PSP and SolO will provide measurements of the total magnetic power and the changes of the spectral break at solar maximum, which will allow us to evaluate the corresponding solar modulation in the near future.

Our results will be important for comparing direct GCR measurements by PSP and SolO in the ecliptic plane. They will also be important for comparing to indirect GCR measurements obtained by observations of the inverse-Compton gamma-ray flux caused by GCR electrons up-scattering solar photons. In that case, it will be possible to probe GCR fluxes far outside the ecliptic plane. It may be that solar modulation at the high latitudes of the Sun is as small as we have predicted for inside the ecliptic plane, in which case the gamma-ray flux will be approximately symmetric around the Sun. It might also be the case that modulation at the high-latitude regions is larger than in the ecliptic plane, in which case the gamma-ray flux will be fainter outside the plane.

Going further, it will be important to assess the impact of coronal magnetic fields on GCR propagation. Here, we have made predictions down to 0.1~au in heliospheric radius, taking into account modulation in the IMF but ignoring coronal magnetic fields. Ultimately, we need predictions down to even smaller heliospheric radii, both to understand the inverse-Compton emission from the solar halo and the emission from the disk caused by hadronic GCR.

As a final note, although our improved force-field model fits the limited data much better than the rigidity-independent model, we emphasize that our model is not complete yet. One missing piece of physics is particle drift. First, because the drift velocity is divergence-free and does not involve any wave-particle scattering, particle drift could speed up or slow down the transport of GCRs toward the Sun, depending on the charge sign of GCRs and the magnetic polarity of the Sun. Second, particle drift also enables GCRs to move between the polar and ecliptic regions of the Sun. Both factors are three-dimensional effects that are beyond the scope of the force-field model. Because our improved force-field model includes only the diffusive effect from charged-particle interactions with magnetic turbulence, the results shown in this work are independent of the charge-sign effect from particle drift. Despite the lack of particle drift, we have shown that the improved force-field model already provides predictions close to the $G_r$ measurements from the Helios and Pioneer missions.

To further quantify the contribution of modulation from particle drift, a comparison between our results and the solutions from the full cosmic-ray transport equation is warranted. This can help distinguish between the charge-sign effect due to particle drift and the diffusive effect due to particle scattering with magnetic turbulence. For this purpose, we encourage inner heliospheric predictions of numerical cosmic-ray transport models from the broader cosmic-ray community.

\vspace{0.2cm}
We are grateful for helpful discussions with Xiaohang Chen, Ilias Cholis, Ofer Cohen, Federico Fraschetti, Joe Giacalone, J\'{o}zsef K\'{o}ta, Mikhail Malkov, Johannes Marquardt, and Igor Moskalenko.
This work was supported by NASA
grants Nos.\ 80NSSC20K1354 and 80NSSC22K0040.
J.F.B. was additionally supported by National Science Foundation grant No.\ PHY-2012955.

\vspace{0.1cm}
\appendix
\section*{GCR Helium nuclei}\label{appendix:helium}

\begin{figure}[tbp] 
   \centering
   \includegraphics[width=0.47\textwidth]{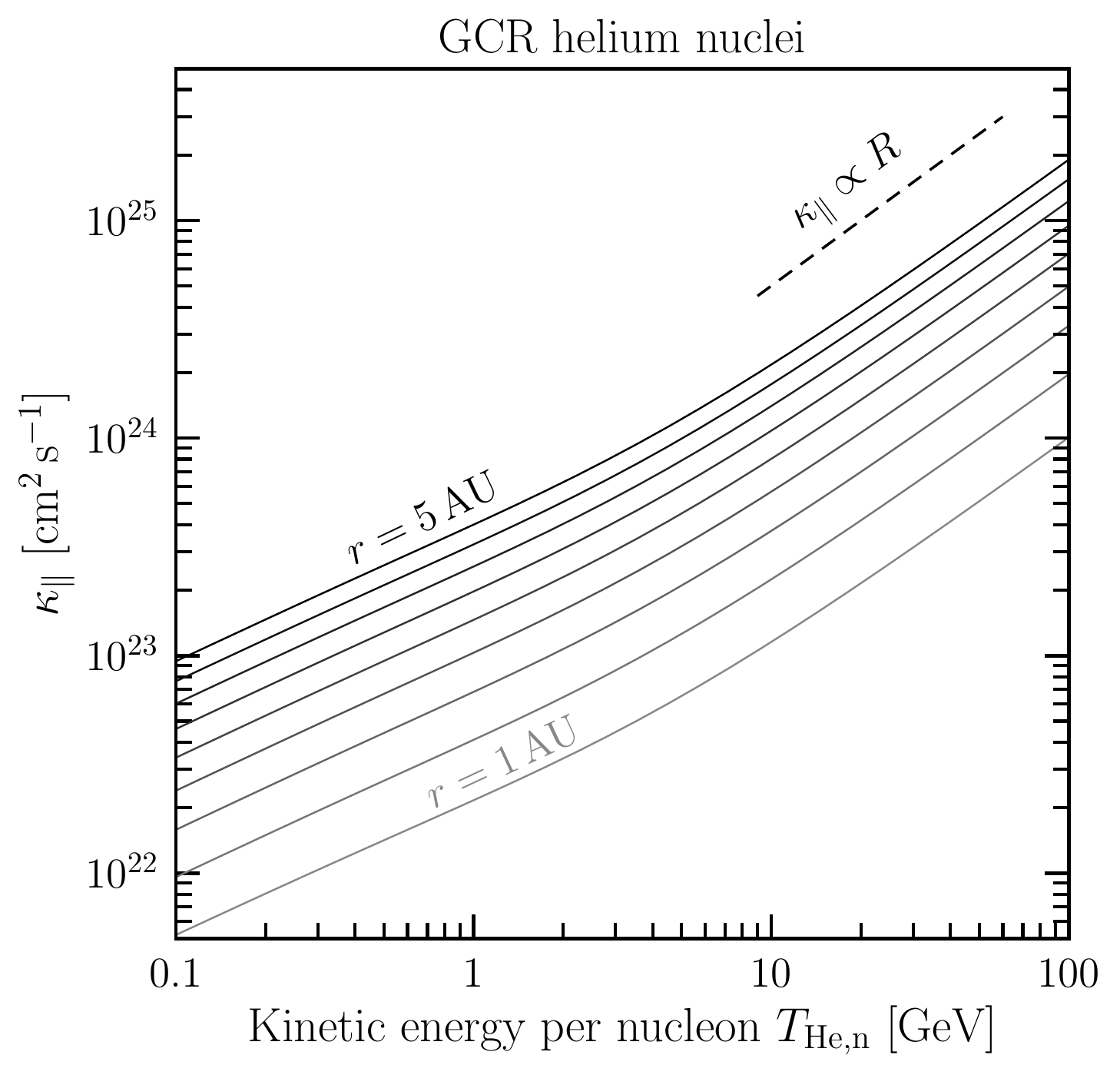}
   \caption{Parallel diffusion coefficient of GCR helium nuclei for heliocentric distances between 1 and 5~au in steps of 0.5~au.}
   \label{fig:kappa_T_helium}
\end{figure}

\begin{figure}[tbp] 
   \centering
   \includegraphics[width=0.47\textwidth]{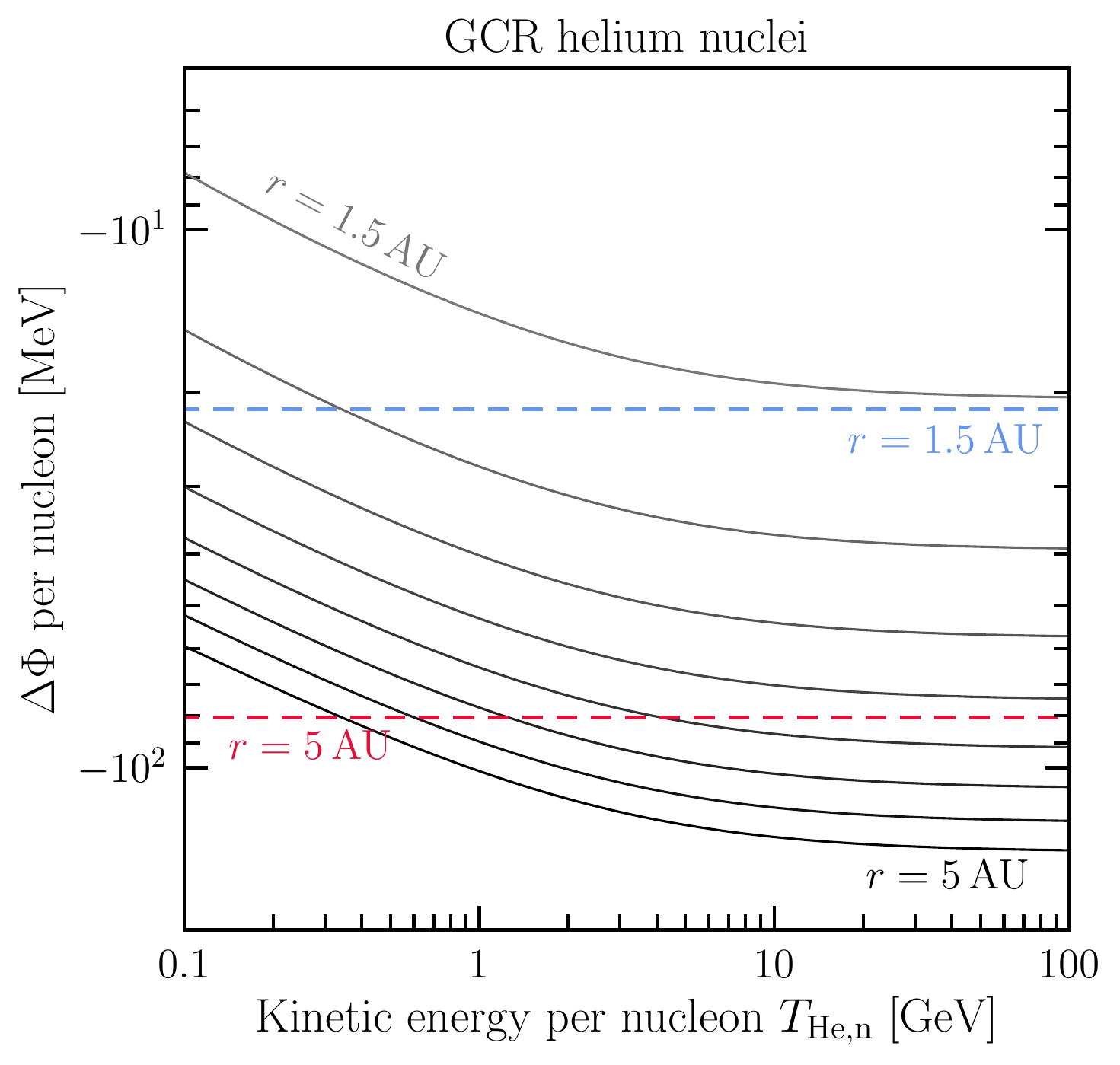}
   \caption{Calculated change in rigidity-dependent $\Phi$ per helium nucleon from $1~{\rm au}$ to radii $r$ between 1.5 and 5~au every 0.5~au. As a comparison, we show the decrease of rigidity-independent $\Phi$ for $r=1.5~{\rm au}$ (top dotted line) and $r=5~{\rm au}$ (bottom dotted line).}
   \label{fig:modulation_potential_helium}
\end{figure}

\begin{figure}[tbp]
   \centering
   \includegraphics[width=0.47\textwidth]{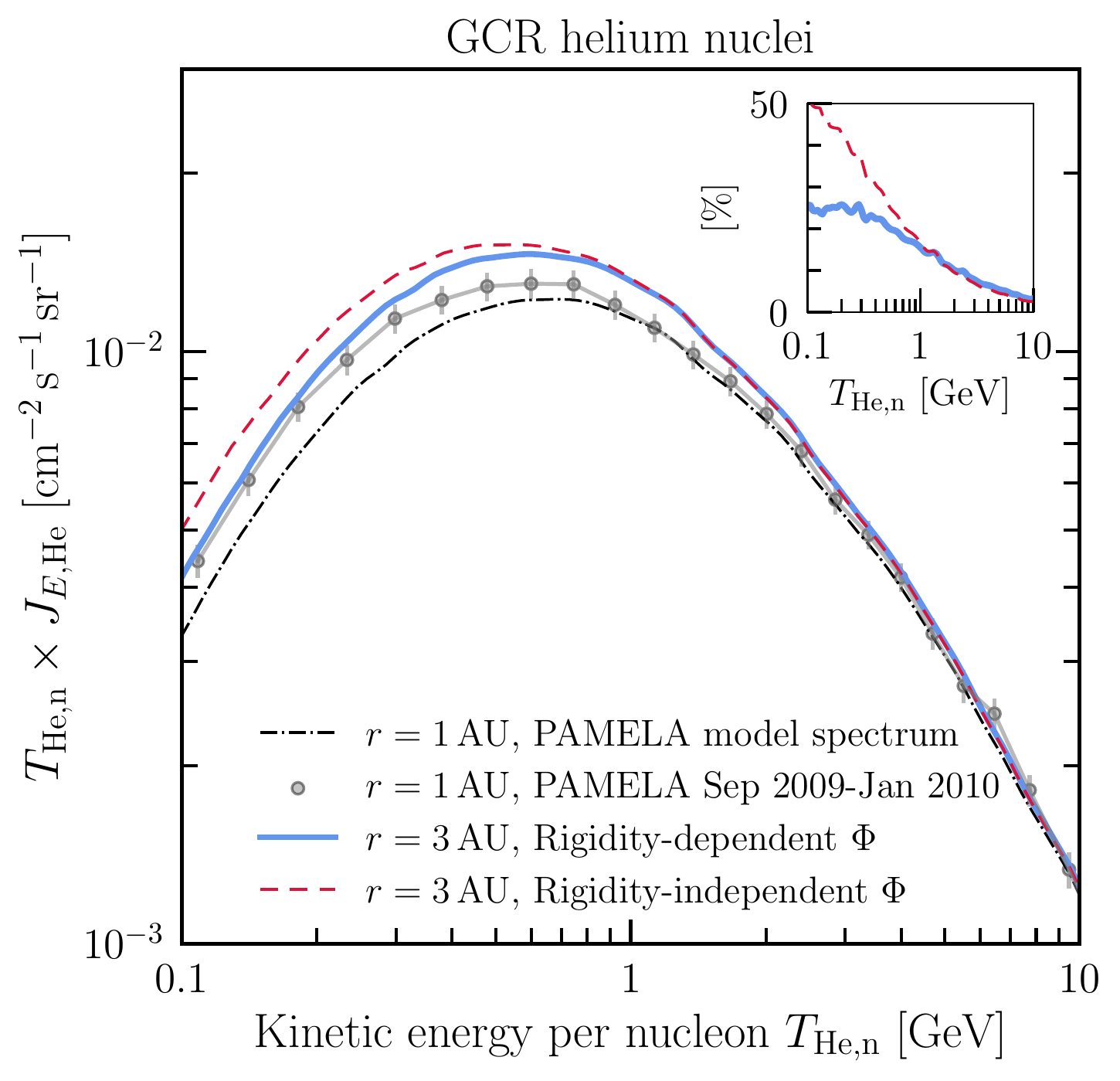}
   \caption{Predicted spectrum of GCR helium nuclei at 3~au. The blue solid line indicates our result. The red dashed line indicates the rigidity-independent $\Phi$ model. The gray circles are the PAMELA helium nuclei observations with the error bars denoting the combined statistical and systematic errors; the black dash--dotted line is their model spectrum~\citep{2020ApJ...893..145M}. Inset: the percentage change of the intensity of GCR helium nuclei from 1--3~au.} 
   \label{fig:CR_flux_helium}
\end{figure}

In this appendix, we show the numerical results of $\kappa_\parallel$, $\Delta \Phi$, and $J_E\left(T,r\right)$ for GCR helium nuclei at $1~{\rm au} < r < 5~{\rm au}$.

Figure~\ref{fig:kappa_T_helium} shows $\kappa_\parallel$ of GCR helium nuclei as a function of kinetic energy per nucleon $T_{\rm He, n}$. Here, we have extrapolated the numerical values of the magnetic PSD and $f_{\rm b}$ from Equations~\eqref{eq:E_B} and~\eqref{eq:frequency_break}, respectively, assuming they are valid at $1~{\rm au} < r < 5~{\rm au}$.

Figure~\ref{fig:modulation_potential_helium} shows $\Delta \Phi $ per nucleon of the GCR helium nuclei (solid lines) from our model. (Note that $\Delta \Phi$, which is defined as $\Phi\left(T, r\right) - \Phi\left(T, 1~{\rm au}\right)$, is negative for $r>1~{\rm au}$ because larger $r$ has smaller $\Phi$.) As a comparison, we also show $\Delta \Phi$ from the rigidity-independent $\Phi$ model in Equation~\eqref{eq:R_indep_compact}, assuming $\earthpotential=800~{\rm MeV}$ per helium nuclei (i.e., $\earthpotential=400~{\rm MeV}$ per proton inside the helium nuclei) and $\eta = {1.1}$.

Figure~\ref{fig:CR_flux_helium} shows the predicted spectrum (blue solid line) of GCR helium nuclei at $r=3\:{\rm au}$ from our model. We use the black dash--dotted line as the input spectrum for $J_E\left(E+\Delta\Phi, {\rm 1~au}\right)$ in the force-field solution in Equation~\eqref{eq:GCR_intensity_02AU}. (We note that below $T_{\rm He, n}<0.8$~GeV, the model spectrum provided in~\cite{2020ApJ...893..145M} is systematically lowered by $10\%$--$15\%$ with respect to the PAMELA measurements. There have not been compelling solutions to resolve this issue. For the simplicity of the analysis, we picked the model line as the input GCR spectrum at 1~au.) We also show with a red dashed line the energy spectrum of GCR helium nuclei at ${\rm 3 \: au}$ from the rigidity-independent $\Phi$ model in Equation~\eqref{eq:R_indep_compact}, assuming $\earthpotential=800~{\rm MeV}$ per helium nuclei and $\eta={1.1}$. We see that our result has a much lower decrease in GCR intensity than that of the rigidity-independent model for $T_{\rm He, n} \lesssim 1~{\rm GeV}$.

\clearpage
\newpage

\bibliography{reference}
\bibliographystyle{aasjournal}

\end{CJK*}
\end{document}